\documentclass[review]{elsarticle}

\usepackage{lineno,hyperref}
\usepackage{graphicx}
\usepackage{color}

\usepackage{float}
\usepackage{hyperref}
\hypersetup{
    colorlinks=true,
    linkcolor=blue,
    filecolor=magenta,      
    urlcolor=cyan,
}

\usepackage[boxruled,vlined,linesnumbered]{algorithm2e}

\usepackage[a4paper, total={6in, 10in}]{geometry}
\modulolinenumbers[5]

\SetKwRepeat{Do}{do}{while}%

\journal{Information and Software Technology Journal}

\begin{document}

\begin{frontmatter}


\title{Handling Constraints in Combinatorial Interaction Testing in the presence of Multi Objective Particle Swarm and Multithreading}

\author{Bestoun S. Ahmed*}

\address{Istituto Dalle Molle di Studi sull’Intelligenza Artificiale (IDSIA), CH-6928 Manno-Lugano, Switzerland and 

Department of Computer Science, Faculty of Electrical Engineering, Czech Technical University, Karlovo n´am. 13, 121 35 Praha 2, Czech Republic

albeybes@fel.cvut.cz}

\author{Luca M. Gambardella}
\address{Istituto Dalle Molle di Studi sull’Intelligenza Artificiale (IDSIA),CH-6928 Manno-Lugano, Switzerland

luca@idsia.ch}

\author{Wasif Afzal}
\address{School of Innovation, Design and Engineering, Mälardalen University, Sweden

wasif.afzal@mdh.se}

\author{Kamal Z. Zamli}
\address{Faculty of Computer Systems and Software Engineering, University Malaysia Pahang, Gambang, Malaysia

kamalz@ump.edu.my}

\begin{abstract}
\linespread{1.1}\selectfont

\textbf{Context}: Combinatorial testing strategies have lately received a lot of attention as a result of their diverse applications. In its simple form, a combinatorial strategy can reduce several input parameters (configurations) of a system into a small set based on their interaction (or combination). In practice, the input configurations of software systems are subjected to constraints, especially in case of highly configurable systems. To implement this feature within a strategy, many difficulties arise for construction. While there are many combinatorial interaction testing strategies nowadays, few of them support constraints.

\textbf{Objective}: This paper presents a new strategy, to construct combinatorial interaction test suites in the presence of constraints.

\textbf{Method}: The design and algorithms are provided in detail. To overcome the multi-judgement criteria for an optimal solution, the multi-objective particle swarm optimisation and multithreading are used. The strategy and its associated algorithms are evaluated extensively using different benchmarks and comparisons. 

\textbf{Results}: Our results are promising as the evaluation results showed the efficiency and performance of each algorithm in the strategy. The benchmarking results also showed that the strategy can generate constrained test suites efficiently as compared to state-of-the-art strategies. 

\textbf{Conclusion}:The proposed strategy can form a new way for constructing of constrained combinatorial interaction test suites. The strategy can form a new and effective base for future implementations. 

\end{abstract}


\end{frontmatter}

%\linenumbers

\section{Introduction}

In the last decade, various studies on combinatorial interaction approaches have gained a lot of awareness and several test generation approaches were developed. In software engineering, combinatorial interaction testing (CIT) aims to generate minimised test suites that manipulate the variables of input parameters based on their combination. Each combination could form a specific configuration of the software-under-test (SUT). The goal is to cover all possible $t$-combinations (sometimes called $t-tuples$) by an optimised set (where $t$ is the interaction strength) \cite{Ahmed2015}. This could be a difficult task in case of highly configurable systems, which leads to combinatorial explosion and non-deterministic polynomial-time hard (NP-hard) problems \citep{Lei2008}. In addition, the problem of constrained interactions has recently appeared \cite{Petke2015}. 

Nowadays, software development is shifted from building an isolated, product-by-product approach to software product lines (SPL) \cite{Kim2011}. In addition to the many features this approach provides like minimising the cost and market reachability time, it facilitates the idea of customisable software products. With this approach, there are many customisable features that could be added to or extracted from the functionality of a specific software based on the needs of the developers or customers. Eclipse represents a well-known example of SPL in which many functions and plug-ins (i.e., having different features) can be added or extracted from its main framework \cite{Michael2011}. Evidence showed that most of the faults may result due to the interaction among these features \cite{Ahmed2015, Ahmed2016}. Hence, all the interactions must be tested carefully. However, in reality, there are many constraints among the features. Some features must or must not appear with others and they may be included or excluded from the test suite during the testing process. CIT strategies can tackle these interactions efficiently, however, with ordinary strategies, it is not possible to satisfy all the constraints and they may contain some invalid combinations in the final constructed test suite.  

There are many approaches to construct CIT test suites in the literature. Among those approaches, evidences revealed that the use of meta-heuristic algorithms could achieve an optimum or near-optimum combinatorial set, covering every possible interaction of input parameters (or functions) \cite{Afzal2009, Nie2011}. Most recently, different meta-heuristic algorithms have been adapted to solve this problem such as Simulated Annealing (SA) \cite{Cohen2003}, Genetic Algorithms (GA) \cite{Shiba2004}, Tabu Search (TS) \cite{Gonzalez-Hernandez2015}, Ant Colony Algorithm (ACA) \cite{Chen2009}, Cuckoo Search (CS) \cite{Ahmed2015}, and many other algorithms. Despite the wide range of approaches and algorithms used in generating the combinatorial interaction set, there is no "universal" strategy that can generate optimised sets for all configurations since this problem is NP-hard problem \cite{Lei1998}. Hence, each strategy could be useful for specific kinds of configurations and applications.

Although different strategies have been developed, the problem of search space complexity is still the same. As mentioned earlier, the main aim of CIT strategies is to cover entire interactions of input parameters by using smallest set. Hence, the strategy needs to search for a combination that can cover much of those interactions. To determine the number of interactions covered, the strategy must search for them among a large number of interactions which will definitely consume the program run time as well as other resources. It will likewise cause the program to take more iteration for searching within the meta-heuristic algorithm. In addition to this problem, nowadays SPLs have many features to customise that leads to many input parameters for CIT strategy. The problem of generating input parameters’ combination represents another serious problem, given consumption of time and resources. This problem appears clearly as input parameters continue to grow in size, since generally, most of the algorithms’ complexities are growing alongside the number of parameters. To overcome this problem, a special algorithm is needed to be combined with efficient data structures in order to speed up the generation and sorting process. In addition to these existing challenges to the implemented strategies, a few of them can satisfy the constraints in the final generated test suite. This will add extra complexity in designing efficient CIT strategies. 

In our earlier research \cite{Ahmed2015, Ahmed2011, Sahib2014}, we have examined Particle Swarm Optimisation (PSO) within a CIT strategy to generate ordinary combinatorial test suites. The strategy has been modified and implemented also for the same purpose by other researchers recently \cite{Wu2015}. In both cases, PSO has outperformed other strategies in different experiments. However, none of these researches are suitable for the constrained CIT. Adding this feature to the strategy will change the nature of the problem and the search space itself. It also changes the fitness function of the optimisation algorithm. In addition, the aforementioned problems within those strategies, are still not solved properly. 

In order to solve these problems and to cope with the practical test generation process, this paper presents a new strategy that tries to generate constrained combinatorial interaction test suites efficiently and provides a new approach for the design and implementation of these strategies. Owing to the nature of constrained combinatorial generation, multi-objective PSO is used within the strategy. Furthermore, the strategy expands our earlier research and gives more practical results depending on the new algorithms that we have designed and implemented. 

The rest of this paper is organised as follows. Section \ref{motivatingExample} illustrates a practical model of the problem as a motivating example, using an SPL case study. Section \ref{CCASection} presents the mathematical notations, definitions, and theories behind the CIT and constrained CIT. Section \ref{RelatedWork} summarises recent related works and reviews the existing literature. Section \ref{MOPSOsection} reviews multi-objective PSO and discusses its features for this research. Section \ref{TheStrategySection} discusses the design concepts of the strategy and its implementation. The section also illustrates how we adapt different algorithms for constrained CIT. Section \ref{EvaluationSection} contains the results of different stages of the evaluation process. Section \ref{DiscussionSection} discusses the experimental results. Section \ref{ThreadsSection} shows some threats to validity in this research.  Finally, Section \ref{ConclusionSection} concludes the paper.

\section{Motivating Example} \label{motivatingExample}

\begin{figure} [h!]
\centering

\includegraphics [width= 0.6\linewidth]{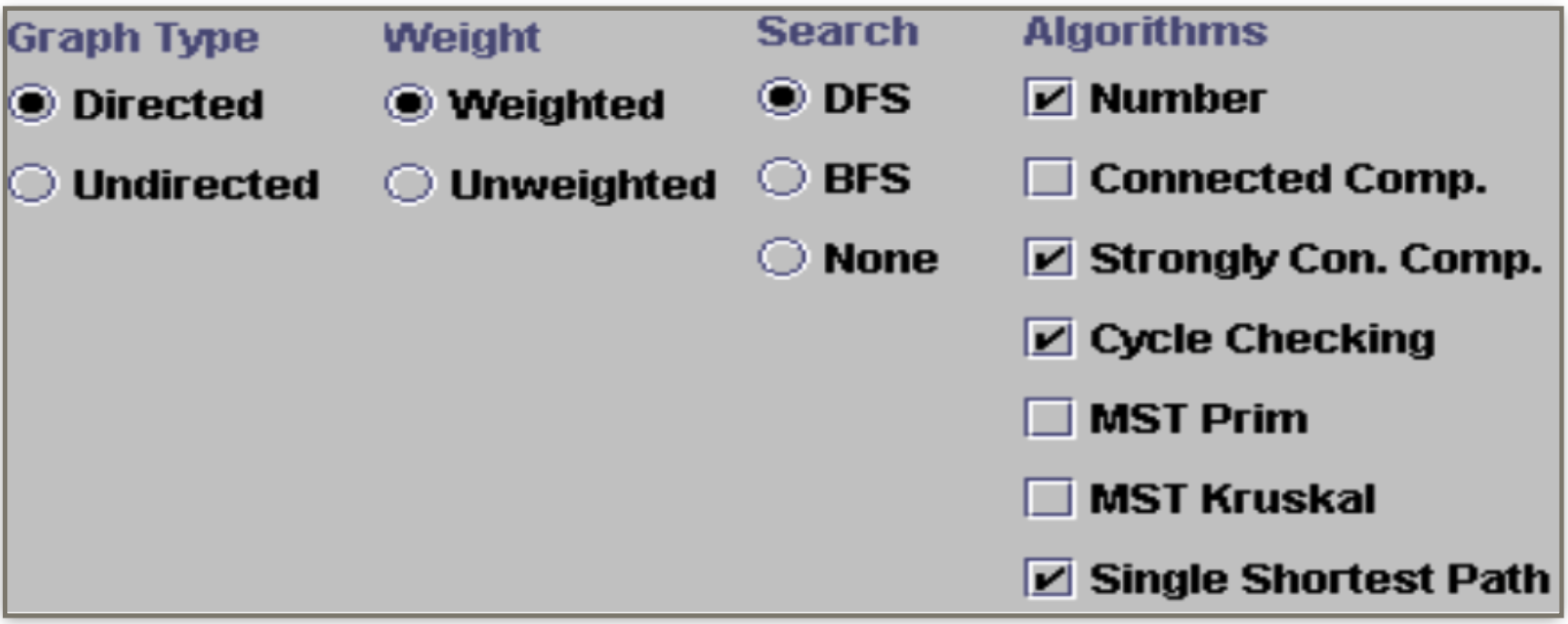}

\caption{GPL specification GUI}

\label{GUIofGPL}
\end{figure}

To illustrate the constrained combinatorial interaction in practice, we adopt a real canonical example from Software Product Lines (SPL) called Graph Product Line (GPL) \cite{Lopez-Herrejon2015}. GPL is a configurable system in which the combinations lead to a product with basic graph algorithms and graph types \cite{Lopez-Herrejon2001}. It is implemented in a way that all the applications that come out as a product do not have the same combination of features. Figure \ref{GUIofGPL} shows a GUI with all specifications of the GPL.

%================================ Figure 1

% ===============================

\begin{figure} [h!]
\centering

\includegraphics [width= 1.0\linewidth]{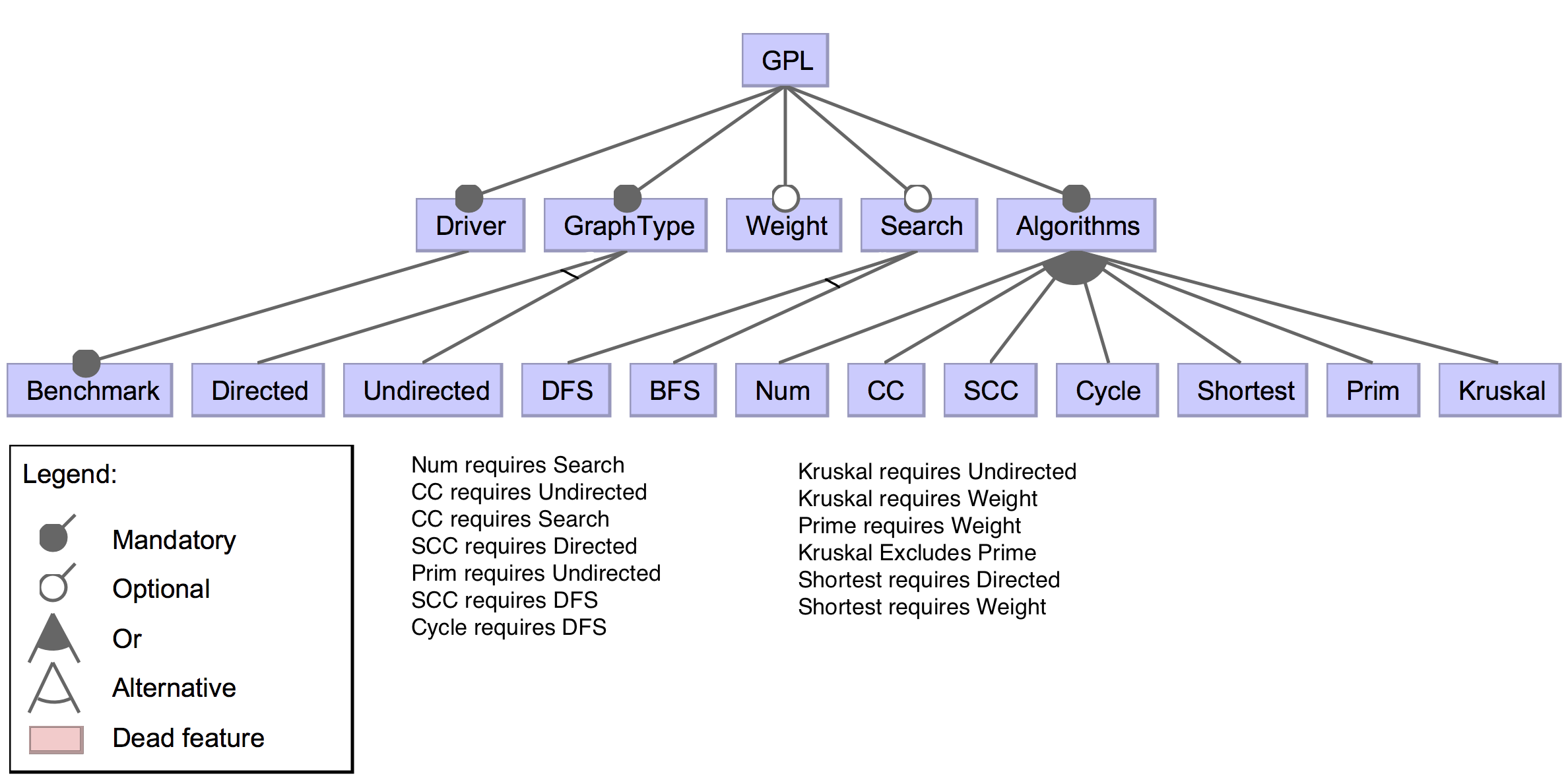}

\caption{The feature model of Graph Product Line (GPL)}

\label{FeatureModel}
\end{figure}

As can be seen in Figure \ref{GUIofGPL}, the produced graph type could be “Directed” or “Undirected,” and the edges of the graph could be “Weighted” or “Unweighted” (i.e., non-negative numbers). In addition to these two features, the graph needs at least one search algorithm which can be a breadth-first search (BFS) or a depth-first search (DFS). Finally, the graph needs one or more of the following algorithms: Vertex Numbering (Number), Connected Components (Connected), Strongly Connected Components (Strongly Connected), Cycle Checking (Cycle), Minimum Spanning Tree (MST Prim, MST Kruskal), and Single-Source Shortest Path (Shortest). Figure \ref{FeatureModel} shows a feature model for these graph features. 

%============================ Figure 2

%============================

The feature model in Figure \ref{FeatureModel} shows a standard model of the illustrated features in Figure \ref{GUIofGPL}. The graph illustrates that the product in the root feature (GPL) has four core functionalities and one driver program (Driver). The driver chooses the example from “Benchmark” feature to apply the graph “Algorithms” to. Each of the functionalities (Graph Type) and (Search) have two alternative features. Finally, one of the following algorithms must be followed with the product: connected components (CC), numbering of nodes in the traversal order (Num), cycle checking (Cycle), strongly connected components (SCC), shortest path (Shortest), Kruskal’s algorithm (Kruskal), or minimum spanning trees with Prim’s algorithm (Prim). 

To apply the CIT method on this feature model, the parameters and values must first be specified. Here, the four functions become parameters and their features become values for these parameters. Table \ref{TableFunctionFeature} illustrates this. 

%=================================== Table 1
\begin{table} [h!]
\centering
\includegraphics [width= 0.8\linewidth]{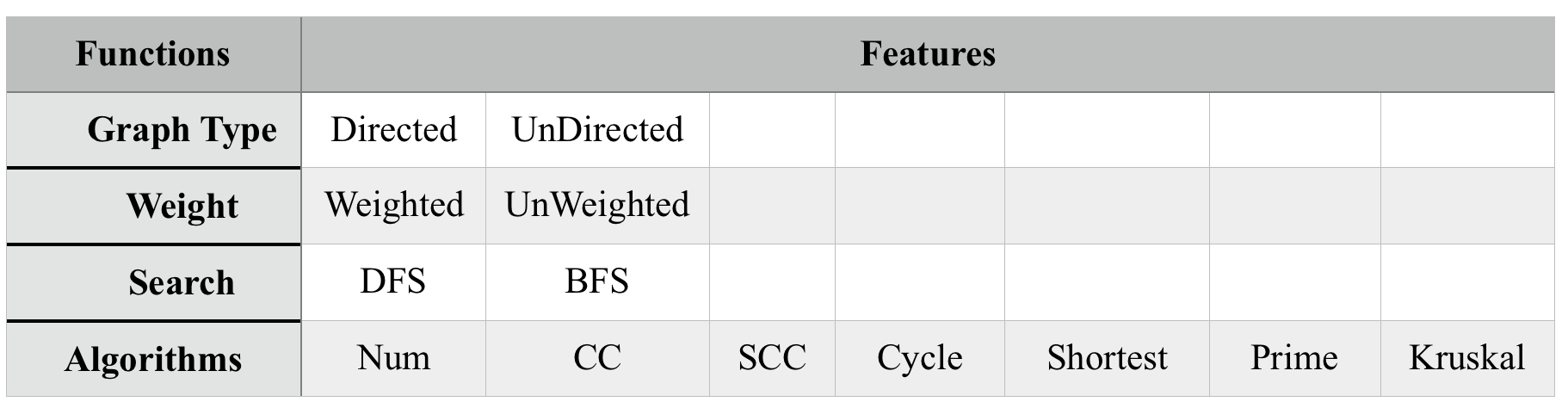}
\caption{Function and feature list of the GPL feature model}
\label{TableFunctionFeature}
\end{table}

% =====================================

The arrangement in Table \ref{TableFunctionFeature} shows clearly that there are $2 \times 2 \times 2 \times 7$ possible feature configurations, which equals 57 possibilities. We can reduce this exhaustive test suite by taking an interaction of the features among the functions. In addition to the reduction advantage, this can tackle the faults caused by the interaction of these features. This idea is supported by much evidence in the literature in which it was shown that faults are likely to occur because of the interaction of few of the system’s parameters. Of course, for this system, the test suite size is not that large. However, most of the current SPL systems have many features that lead to a huge size of the test suites. This, in turn, makes them impossible to run in reality. 

Using the CIT approach, the test cases could be reduced dramatically. For example, if we take the interaction of the features within two functions, i.e., when the interaction strength $t = 2$, then we can cover all the interactions with only 14 test cases. This great reduction in the number of test cases could be useful when there are no constraints among the features. In other words, the output test suite may contain invalid combinations when we have constraints in the system. In the case of GPL, we note from the feature model in Figure \ref{FeatureModel} that there are many constraints that must be satisfied, as illustrated at the bottom of the graph. For instance, one of the products that emerges from the combinatorial test suite is (GPL : Driver . Directed . Unweighted . BFS . SCC). However, this test case does not reveal the right product because as can be seen in the constrained list in Figure \ref{FeatureModel}, the algorithm type SCC requires the search type to be DFS, while in the generated test, it is BFS. Hence, this test case is not valid as it violates the constraint. 

This shows that the ordinary CIT strategy could not be used for such situation. Constrained CIT strategy is an alternative way of generating combinatorial interaction test suites that satisfies the provided constraints. For example, Table \ref{GPLTestSuite} could be a test suite for possible products. Note that all constraints are satisfied in the table. The ticked cells indicate that the feature is available.

%=================================== Table 2
\begin{table} [h!]
\centering
\includegraphics [width= 1.0 \linewidth]{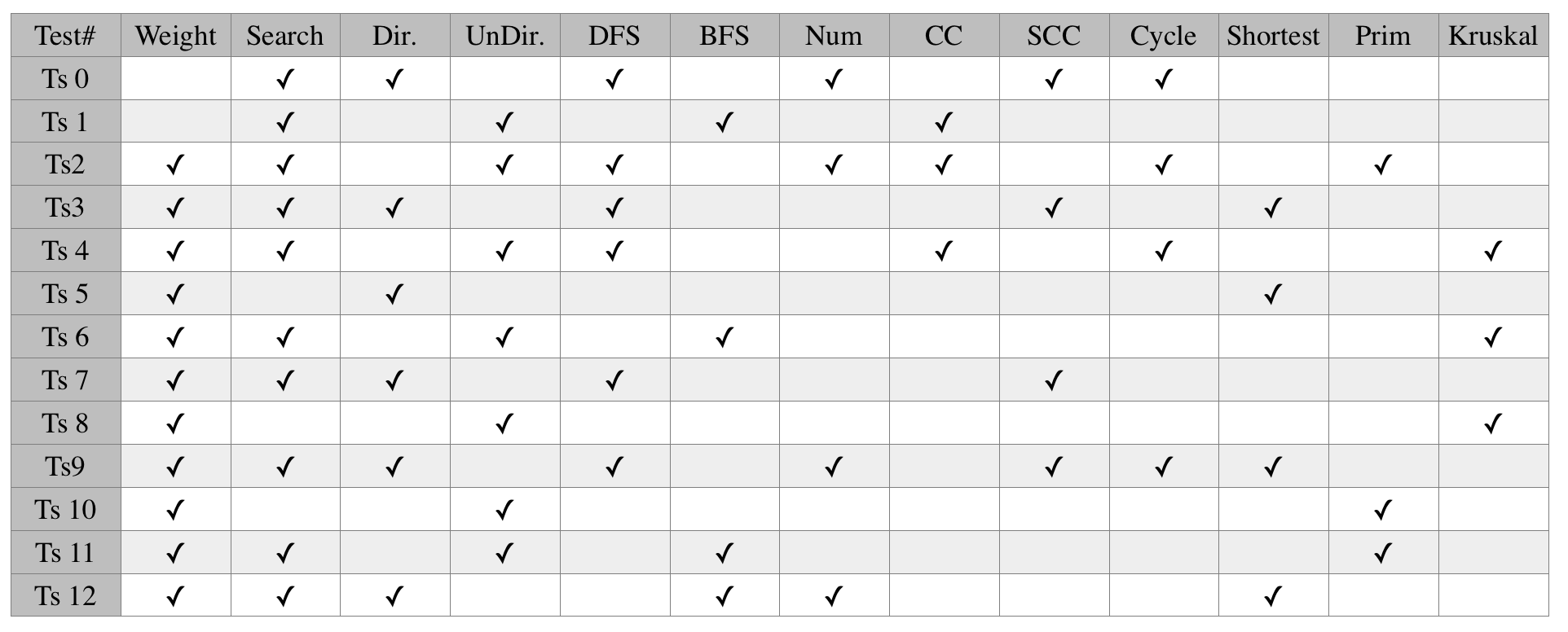}
\caption{Possible test suite for GPL}
\label{GPLTestSuite}
\end{table}
% =====================================

\section{Constrained Covering Array (CCA)}\label{CCASection}

Covering array (CA) is a combinatorial mathematical object that represents the optimised set of combinatorial test suites. The object assumes that all interactions among the features are occurring within one array. The origin of CA is essentially from another mathematical object called orthogonal array (OA) \cite{Walter1986}.
\\

\textbf{Definition 3.1 :} An $OA_\lambda (N; t, k, v)$ is an $N \times k$ array, where for every $N \times t$ sub-array, each $t-tuple$ occurs exactly $\lambda$ times, where $\lambda = N/v^t$; $t$ is the combination strength; $k$ is the number of input functions ($k \geq t$); and $v$ is the number of features or levels associated with each input function. 
\\

For example, we can see that an $OA (9; 2, 4, 3)$ can cover two interactions of four input functions when these functions have three features each with only nine rows in the array. The exact repetition of the interactions eliminates the use of OA when the number of features is growing. To make OA more flexible, the CA notation is introduced as an alternative. CA is suitable when the number of functions and features are increasing \cite{Kacker2013}, which is defined as bellow \cite{Colbourn2010}: 
\\

\textbf{Definition 3.2 : }In its general form,  $CA_\lambda(N ; t, k, v)$ is an $N \times k$ array over ($0, . . . , v-1$) such that every $B=\lbrace b_0, ..., b_{d-1}\rbrace  \in$ is $\lambda -covered$ and every $N \times d$ sub-array contains all ordered subsets from v features of size t at least $\lambda$ times, where the set of column $B=\lbrace b_0, ..., b_{d-1}\rbrace \supseteq \lbrace 0, ..., k-1\rbrace$. 
\\

As far as we are looking for the optimal array, the value of $\lambda$=1 which means that all the $t-tuples$ of the input functions occur at least once and thus the notation becomes $CA(N;t,k,v)$ \cite{Hartman2004}. However, this notation assumes that the number of features in each function is equal, which is not the case in real applications. Normally each function has a different number of features depending on the system-under-test. To this end, mixed covering array (MCA) is used as a practical alternative, which is defined as below \cite{Gonzalez-Hernandez2015}:
\\

\textbf{Definition 3.3 :} $MCA (N; t, k, (v_1, v_2, ... v_k))$, is an $N \times k$ array on $v$ levels, where the rows of each $N \times t$ sub-array cover and all t interactions of values from the t columns occur at least once . 
\\

For more flexibility in the notation, the array can be presented by $MCA (N;d, v^k)$ and can be used for a fixed-level CA, such as $CA (N;t, v^k)$  \cite{Gonzalez-Hernandez2015}. So far, these notations have been used within CIT literature. However, as illustrated previously, these mathematical objects may not suit the requirements of current applications. Modern software systems are subjected to constraints. There would be many combinations of features that must or must not appear in the array. Owing to such constraint, many of the rows in the CA or MCA would be invalid for testing. The following definitions are derived from the previous definitions \cite{Cohen2008}: 
\\

\textbf{Definition 3.4 : }A Constrained CA, donated as $CCA (N;t,k,v)$, is an $N \times k$ sub-array with $v$ features that have constraints $C$ in which every $N \times t$ sub-array contains constraints of all orders subset of size $t$ at least once with $v$ features. 
\\

In the same way, a more flexible notation to show the mixed number of features among the input functions is to use constrained MCA as in the following definition:  
\\

\textbf{Definition 3.5 : }A Constrained MCA donated as $CMCA (N; t, k, (v_1, v_2, ... v_k))$ is an $N \times k$ sub-array with $v$ features that have constraints C in which every $N \times t$ sub-array contains constraints of all orders subset of size $t$ at least once with $v$ features.

\section{Review of the literature and related works} \label{RelatedWork}

As described previously, CIT aims to cover every $t-tuple$ of any input interaction of a SUT at least once \cite{Bryce2005}. These interactions are captured by constructing the combinatorial object known as CA. Kuliamin and Petukhov \cite{Kuliamin2011} present a survey on the methods of constructing CAs. Inline with that research and other perspectives in the literature in general, the methods for constructing CAs can be categorised into three categories \cite{Kuhn2013, Cohen2007}: (1) algebraic methods (2) meta-heuristic methods and (3) greedy search methods. Algebraic methods use extremely fast mathematical techniques (both direct and recursive) \cite{Cohen2008} but their applicability is limited to certain special combinatorial test structures \cite{Kuhn2013}. Examples of using algebraic methods for CA construction include those cited in \cite{Sloane1993, Hartman2005}. Meta-heuristic methods apply complex and iterative heuristic methods that include simulated annealing, tabu search, genetic algorithms, particle swarm and others. Although being computationally intensive, meta-heuristic methods have produced some CAs of the smallest size known \cite{Kuhn2013}. Examples of using meta-heuristic methods includes those cited in \cite{Ahmed2015, Nurmela2004, Cohen2006}. Greedy search methods are known to be faster than meta-heuristic search and are applicable to arbitrary test structures but may or may not produce smallest-size CAs \cite{Nie2011, Kuhn2013}. Examples of using greedy methods include those cited in \cite{Forbes2008, Lei2008}. The three methods of CA generation are sometimes used in combination also. Examples of such integrated approaches include those cited in \cite{Cohen2003, Bryce2007, Myra2008}. 

One of the practical considerations for CA construction includes constraint handling. Certain combinations of parameter values are invalid for testing, e.g., when testing if a web page is displayed correctly in different web browsers running on different operating systems, no test should contain the combination of Internet Explorer and MacOS \cite{Kuhn2013}. Thus, such combinations must be excluded from a test set. Kuhn et al. \cite{Kuhn2013} describe two types of constraints: environment constraints and system constraints. Environment constraints are imposed by the runtime environment of the SUT. For example, the combination Internet Explorer and Linux violates the environment constraint. A system constraint is imposed by the semantics of the SUT \cite{Kuhn2013}. For example, an Internet Banking Application may impose a constraint that a person cannot transfer money exceeding a certain amount. While combinations violating environment constraints must be removed from the test set, those violating system constraints still can be made part of the test set to assess if the system successfully rejects such combinations (e.g., for the case of robustness testing). 

Cohen et al. \cite{Cohen2008, Cohen2007} have summarised constraint handling in a variety of algorithms/tools for constructing CAs and argue that the solutions for handling constraints in CIT are less than satisfactory. Zhong et al \cite{Zhong2016} have combined constraint-based generation of structurally complex tests (using Korat \cite{Boyapati2002}) with combinatorial testing (using ACTS \cite{RickKuhn2009}). Korat is used to solve structural constraints and generates structures with respect to structural invariants. These structures are then populated with data values using ACTS. The results showed that their approach was able to find a number of previously unknown bugs in the application under test. Hervieu et al. \cite{Hervieu2016} have used a time-aware minimization algorithm based on constraint programming for pairwise coverage of test configurations from feature models. They compared their approach on 224 feature models with two other approaches based on greedy algorithm: SPLCAT \cite{Johansen2012} and MosoPolite \cite{Oster2010}. The results showed that their approach generated up to 60 \% fewer test configurations than the two approaches for 79\% of feature models. According to Bryce and Colbourn \cite{Bryce2006}, asking if a configuration exists that satisfies a set of given constraints is an NP-hard problem. It would then seem fitting to apply meta-heuristic search techniques to handle constraints, however, Cohen et al. \cite{Cohen2007} found that the presence of constraints dramatically increases the time required for constructing CIT samples. Garvin and colleagues \cite{Garvin2009, Garvin2011}have further experimented with algorithmic changes to the simulated annealing algorithm where they report efficiency improvements. However, the experimentation on the use of meta-heuristic search techniques for generating CIT samples in the presence of constraints is still an open research problem. In particular, there is little or no research exploring the use of multi-objective meta-heuristic search techniques for such problem. Given that there are competing constraints such as coverage, total computation time, environmental and system constraints, the use of multi-objective meta-heuristic search techniques can potentially produce competitive results and thus invites evaluation and experimentation. 

In our previous work, we have used PSO for CIT successfully. The method was improved more recently by Wu et al. \cite{Wu2015} for more efficient generation of combinatorial test suites. Within both research studies, PSO is able to generate better test suites in terms of size and outperforms other meta-heuristic strategies. However, using these strategies is not suitable for constrained test suite generation due to the different nature of the problem itself. Cohen et al. \cite{Cohen2007} and Garvin et al. \cite{Garvin2011} developed the AETG, mATG, and SA based strategies to support the generation of constrained test suites by adding a constrained satisfiability solver like the SAT solver. However, evidence showed that adding a solver to the strategy will make the processing and satisfaction of constraints more complicated.  Our new approach is to use multi-objective optimisation to support the generation of constrained test suites for CIT. Owing to its efficiency within CIT strategies and our experience with PSO, we have used multi-objective PSO (MOPSO). The next sections show in detail how MOPSO was used and adapted within our strategies.

\

\

\

\

\section{Multi-objective Particle Swarm (MOPSO)} \label{MOPSOsection}

Generally, an optimisation problem with multi-objective $M$ can be represented mathematically as \cite{Tripathi2007}:

\begin{equation}
\left. \begin{array}{l}
Minimize :  \ \ \ \ \ \ \ \ \ \ \ \ \ \ \ \ \ \ \ \ \ \ \
\overrightarrow f (\overrightarrow x ) = \left[ {f{}_i(\overrightarrow x ),\quad i = 1, \ldots ,M} \right]\\
subject \ to \ the \ constraints : \ \ \ {g_i}(\overrightarrow x )\; \le \;0,\quad j = 1,2, \ldots ,J,\\
\ \ \ \ \ \ \ \ \ \ \ \ \ \ \ \ \ \ \ \ \ \ \ \ \ \ \ \ \ \ \ \  \ \ \ \ \ \ \ \
{h_k}(\overrightarrow x ) = 0,\;\;\;k = 1,2, \ldots ,K,\;\quad 
\end{array} \right\}
\end{equation}

giving that $\mathop x\limits^ \to   = \left[ {{x_1},{x_2},...,{x_d}} \right]$ where $d$ represents space of the decision variable. ${f{}_i(\overrightarrow x )}$ is the $i^{th}$ objective function, ${g_i}(\overrightarrow x )$ is the $j^{th}$ inequality constraint and ${h_k}(\overrightarrow x ) $ is the $k^{th}$ equality constraint. Here the aim is to find an $\overrightarrow x 
$ such that $\overrightarrow f (\overrightarrow x )
$ is optimised. 
In contrast to the single objective optimisation, the Pareto dominance is defined and formulated by Vilfredo Pareto to be used with the multi-objective optimisation as follows \cite{Coello2006}:

A vector $\overrightarrow u  = ({u_1},{u_2}, \ldots ,{u_M})$ is said to dominate a vector $\overrightarrow v  = ({v_1},{v_2}, \ldots ,{v_M})$ or a multi-objective minimization problem, if and only if: 

\begin{equation} 
\forall i \in \left\{ {1, \ldots ,M} \right\},\quad {u_i} \le {v_i} \wedge \exists i \in \left\{ {1, \ldots ,M} \right\}:{u_i} < {v_i}
\end{equation}

where M represents the dimension of the objective space.

A Pareto solution $\overrightarrow u \in U$ is an optimal solution if and only if there is no other solution exist $\overrightarrow v \in U$ such that $\overrightarrow u $ (which is called non-dominated solutions) is dominated by $\overrightarrow v $. A set of these $\overrightarrow u $ forms the Pareto-Optimal Set.

This general definition has been used within different set of optimisation and evolutionary algorithms in different forms to deal with the multi-objective optimisation problems. The Multi-objective Particle Swarm Optimisation (MOPSO) is one of these successful implementation to solve multi-objective problems.

Originally, the Particle swarm optimisation (PSO) is a well-known, behaviour-oriented stochastic optimisation algorithm. The algorithm tries to find the best solution in a solution space based on social behaviour \cite{Clerc2002} . Each individual or possible solution is called a particle and it makes up a part of the solution space which is called swarm. The algorithm takes advantage of the information exchange among these particles to gradually converge towards the best solution by deriving new solutions and adjusting its trajectory towards its previous best position. This feature prevents PSO from having operators like cross-over and mutation, as in a genetic algorithm (GA) or other algorithms, which are reliant on natural evolution to form new solutions \cite{AlRashidi2009} . In the beginning, the algorithm searches in the global space by taking all the particles in the swarm as a neighbourhood and sharing information among the particles. The particle learns the experience from other counterpart particles when they exploit information. This will find a promising region (i.e., a local region) in the global space that the algorithm investigates more to get an improved solution at the end \cite{Zhang2016}. Due to the different natures of the applications since its first emergence until now, PSO has undergone different enhancements and developments; however, the main steps of the algorithm remains the same. Algorithm \ref{PSOGeneralAlgo} shows the main steps of the PSO algorithm.

% ============================= Algorithm 1

\begin{algorithm}
\linespread{1.1}\selectfont

Initialize particles population
\Do{Max iteration is not reached or a stop criterion is not satisfied}{

\For {each particle $p$ with position $xp$}{
calculate fitness value f(xp)
      
      \If{$f(xp)$ is better than $pbest$ }{
      
      $pbest \leftarrow xp$
      
      }
      
      }
      
Define $gbest$ as the best position found so far by any of $p$’s neighbours       
      
      \For {each particle $p$}{
      $vp \leftarrow $ compute velocity($xp$, $pbest$, $gbest$)
      
      $xp \leftarrow $ update position($xp$, $vp$)
      
      }
      
    }

 \caption{PSO general algorithm}

\label{PSOGeneralAlgo}
\end{algorithm}

% =================================

As shown in Algorithm \ref{PSOGeneralAlgo}, the procedure starts by generating a random search space, then evaluates each particle in this space based on the fitness function. The algorithm then chooses the particle to be the “best.” Based on this best particle, the whole search space is updated by adjusting the velocity of movement towards the best solution. This velocity determines the step length of each particle towards its new position.This update function is performed during each iteration. If we suppose that the $i^{th}$ particle in the position $X_i$, is the best particle and we have an N-dimentional search space, the particle position is updated according to the following equations \cite{AlRashidi2009}:

\begin{equation} \label{eq:1}
V_i^{k + 1} = wV_i^{k} + {c_1}{r^k}_{{i_1}}(P_i^k - X_i^k) + {c_2}{r^k}_{{i_2}}(P_g^k - X_i^k)
\end{equation}
\begin{equation} \label{eq:2}
X_i^{k + 1} = X_i^k + V_i^{k + 1}
\end{equation}

where $i = (1,2,…., m)$ and $m$ is the swarm dimension size; $w$ is the inertia weight, $r_{i_1}$ and $r_{i_2}$ are random numbers distributed uniformly between 0 and 1; $c_1$ and $c_2$ are two positive constant positive numbers called the cognitive and social parameter, respectively, $P_i$ is the best position particle “i” achieved based on its own experience, $P_g$ is  the  best  particle  position  based on  overall  swarm’s  experience, and $V$ is the velocity of the particle. As can be seen from equation \ref{eq:1}, the new velocity is found for the $i^{th}$ particle then this velocity is added to current position of the particle in equation \ref{eq:2} which moves  the particle to its new position.

The best particle is judged based on a fitness or objective function in which it determines how far the particle is from the optimal solution. Hence, the aim for each iteration is to get a better value of the objective function by each movement of the particle. However, in reality, there could be two or more objective functions that are conflicting with each other and need to be considered simultaneously to optimise a particle. This in turn leads to the idea of multiple criteria or multi-objective (MO) optimisation.

As mentioned earlier, the idea of multi-objective optimisation has been applied successfully in many research areas \cite{Deb2005}. In both single and multi-objective optimisation, the methodology of optimisation is the same; however, in the multi-objective, there would be a set of alternative solutions that are called “Pareto Optimal” that holds different multiple solutions with trade-offs among the objective functions \cite{Sahib2016}. Adapting this mechanism for an application represents a challenging task due to the different nature of applications. In this research, we used the idea of multi-objective PSO (MOPSO) to generate optimum or near-optimum constrained combinatorial interaction test suite. To adapt this algorithm for constrained CIT, careful designs and different supporting algorithms are needed. We adapt the algorithm successfully and also modified some concepts of the algorithm during the design. The Pareto mechanism in the MOPSO is used in a different form to utilize its use for the constraint CIT. We show these designs and algorithms in detail in the coming sections.

\section{The Strategy} \label{TheStrategySection}

This strategy supports the generation of constrained combinatorial test suites. The following sections show how the strategy designed and implemented. In addition, the implemented algorithms are shown in detail.

\subsection{General Concepts}

The strategy is a combination of different algorithms that have been implemented in .Net environment. Each algorithm solves a specific problem. Here, although the constrained CIT is different from the non-constrained CIT mechanism, there are some similarities in solving both problems. These problems can be categorised into two groups. The first is the legacy problem that originates from CIT strategies such as the input parameter combination generation, search space complexity, and search performance. Second is the constrained problem itself, such as constraint satisfaction.

Although different CIT strategies have been developed, the problem of search space complexity is still the same. As mentioned earlier, the main aim of the combinatorial strategies is to cover the entire interaction of input parameters by the smallest set. Hence, the strategy needs to search for a combination that can cover many of those interactions. To determine the number of interactions covered, the strategy must search for them among a large number of interactions, which will definitely consume program’s time as well as computer resources. It will likewise cause the program to take more iteration for searching within the meta-heuristic algorithm.

In addition to the aforementioned issues, the problem of generating an input parameter combination represents another serious problem. This problem appears clearly as the input parameter continues to grow in size since most of the algorithms’ complexities grow with the number of parameters. To overcome this problem, a special algorithm is needed to be combined with efficient data structures, in order to speed up the generation and the sorting process. Each one of these algorithms is illustrated in detail in the coming subsections. 

\begin{figure} [h!]
\centering

\includegraphics [width= 0.7\linewidth]{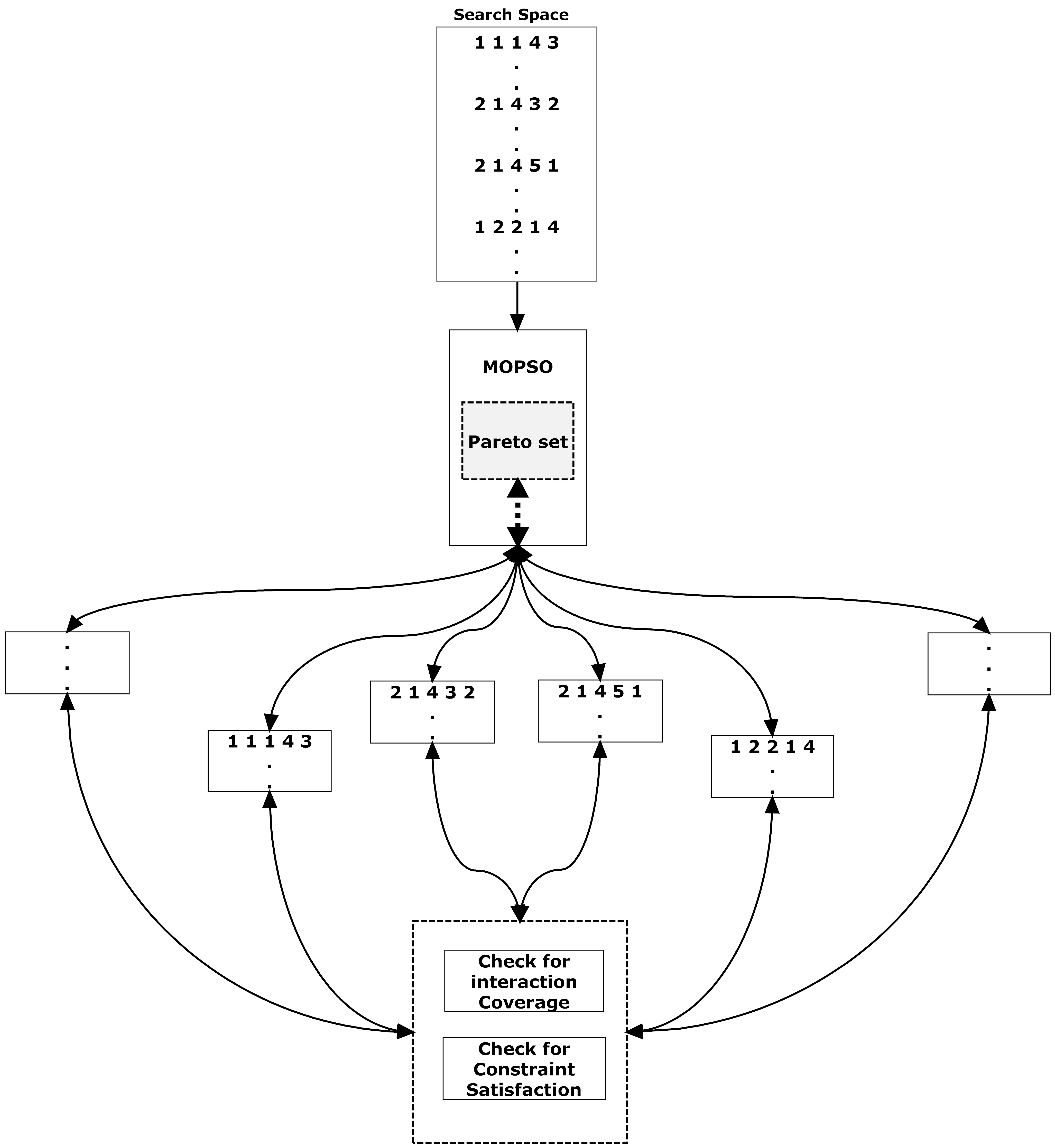}

\caption{The general structure of the strategy}

\label{OctopusStructure}
\end{figure}

The strategy relies on multi-threads in which each thread searches for the covered interactions and also tries to satisfy the constraints. These threads are initialised from a main function. This function keeps monitoring these threads and the results are returned back to it. Each thread has the ability to decide on which solution is better to be returned back to the main function. To achieve this, the search space is divided equally among the threads and the best solution returned back to the main function to form the Pareto set. MOPSO is used to optimise and choose the best solution in the function. The following subsections elaborate on these algorithms in detail. Figure \ref{OctopusStructure} shows the structure of the strategy. 

%============================ Figure 4

%============================

\subsection{The Generation of Input Parameter Combination} \label{InputParameterGenerationSection}

The algorithm uses the CA notation as the base of input with an input configuration file. As shown in Algorithm \ref{ParameterCombAlgo}, the algorithm takes $k$ input parameters and produces t-combinations of them each time adding the combinations to a final array containing all t-combinations of $k$. To generate the $t-tuples$ list, the generation of t-combinations of $k$ is essential.  It should be mentioned here that, the t-combinations generation of $k$ can be done in its ordinary enumeration algorithm, in which all n-bits of the $k$ input parameters must be generated first then filtered base on the specified interaction strength $t$. However, this algorithm is useful for small and medium size of input parameters. As illustrated in the motivation example (Section \ref{motivatingExample}), in practice there could be large number of input parameter as in case of SPL. The enumeration algorithm complexity is approximately $O(n^t)$, where $n$ is the input parameters. Hence, there could be long generation time in case of input parameter and $t$ growing.

\begin{figure} [h!]
\centering

\includegraphics [width= 0.8\linewidth]{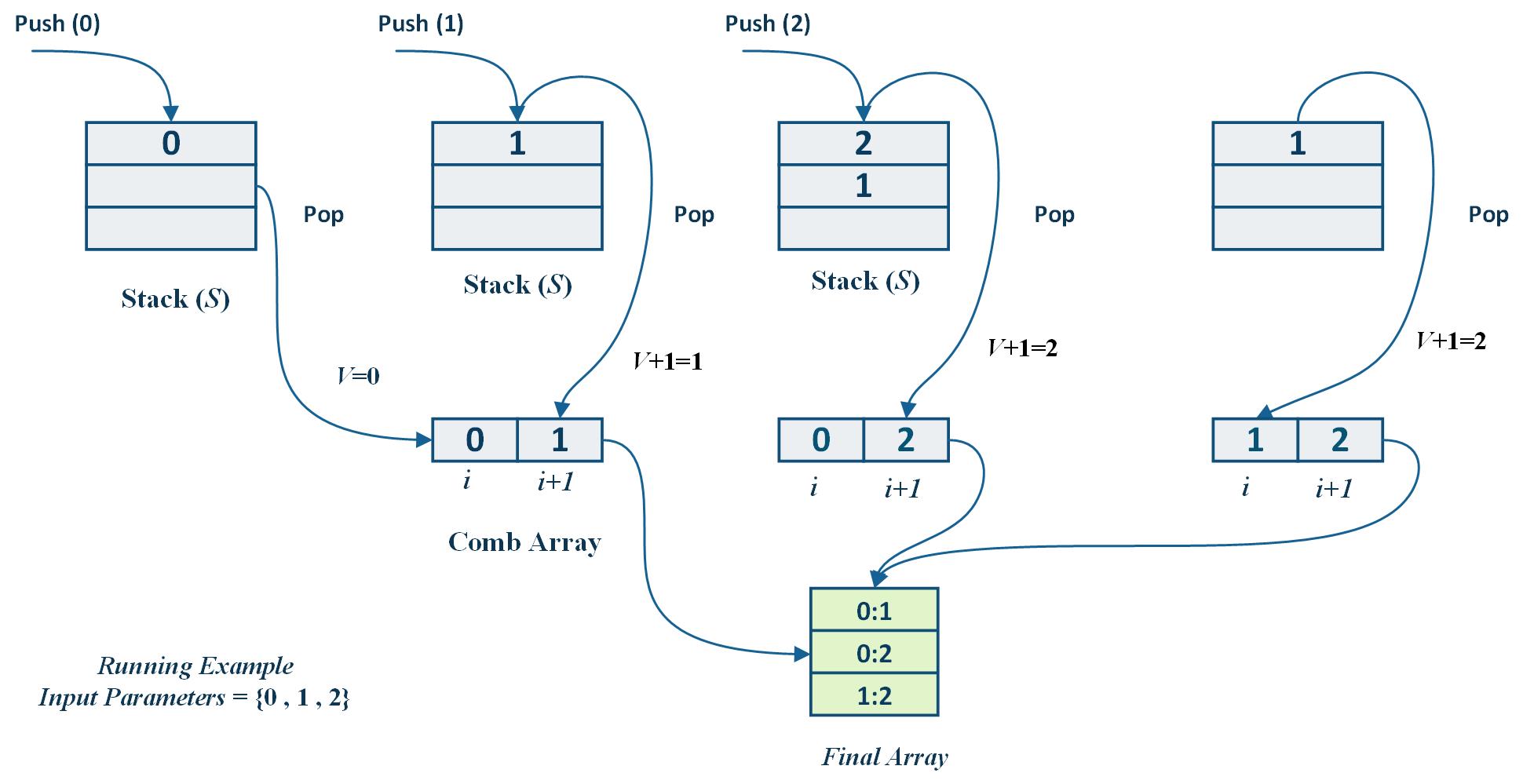}

\caption{A Running Example}

\label{HashTableExample}
\end{figure}

Instead of enumerating all n-bits, we used stack data structure  to hold the parameters permanently by “pushing” them into the stack and then “popping” them when needed during the iterations. Additionally, a temporary array was created with index $i$ to help the generated combinations in each iteration (Steps 1-2). A stack data structure ($S$) was created and the first parameter (0) was pushed inside (Steps 3-4). The algorithm continued to iterate until the stack became empty (Step 5). The index number $i$ of the Comb array was set to length of $S-1$ and the value $v$ of this index $i$ was set to the top value in the stack (i.e. pop) until $v$ was less than $k$ (Steps 6-9). Furthermore, the algorithm continued to increment $i$ and $v$ then puts the value of $v$ into $S$ until the index number was equal to the length of the required interaction strength $t$ (Steps 9-15). The pseudo code is shown in Algorithm \ref{ParameterCombAlgo}. Additionally, for better understanding of the algorithm, a running example is illustrated in Figure \ref{HashTableExample}.

\begin{algorithm}
\linespread{1.1}\selectfont

 \KwIn{Input-parameters $k$ and combination strength $t$}
 \KwOut{All $t$-combinations of k where $k={k_1 , k_2 , k_3 , … k_n}$ }
 Let Comb be an array of length $t$\;
 Let $i$ be the index of Comb array\;
 Create a stack $S$\;
 $S \leftarrow 0$\;
 \While {$S\neq null$}{
 
$i$ =(the length of $S -1$)\;
$v$ = pop the stack value\;

\While {pop value $< k$}{
set Comb of index ($i$) to $v$\;
$i\leftarrow i + 1$\;
$ v\leftarrow v + 1$\;
push $v$ to stack\;

\If{$i=t$}{
Add Comb to final array\;
break\;
}
}
}

 \caption{Parameter Combination Generator}
 \label{ParameterCombAlgo}
\end{algorithm}

Figure \ref{HashTableExample} shows a running example to illustrate how the combinations of input parameters were generated using three input parameters [0, 1, and 2]. With the first parameter pushed into the stack at the start, the algorithm iterated and the stack popped its last value to the $i+1$ index of the Comb array. In the next iteration, the stack pushed by $v+1$ value. The algorithm stopped when the stack became empty. The final array then contained all the interactions of input parameters which are, [(0:1), (0:2), (1:2)].

%============================ Figure 6

%============================

As could be seen in Figure \ref{HashTableExample}, the algorithm kept the previous value of $v$ for the next iteration unless it became greater than the $t$ value. For example, $v = 0$ in the first iteration and in the next iteration, it became $v+1$ which equals to 1. Then it was incremented and pushed into the stack again.

We can use this algorithm for generating combinations for small, medium and large parameter size due to its simpler algorithm structure as compare to the enumeration algorithm. As we note later in the evaluation section, the algorithm has a logarithmic complexity and could perform well for generation of different sizes of input parameters.

\subsection{Searching for the Interaction Coverage}  \label{SearchInTheHashTableSection}

The combinatorial search strategies need to generate all the possible interaction elements between the input parameters, which we called $t-tuples$. This step is vital so as to verify how many of these elements can be covered by the suggested solution. Most of the time, this will be an objective function of the meta-heuristic used in the strategy. It is not clear in most of the implemented strategies which data structure and searching mechanism they used since they are close-sourced. However, for the known strategies, there are different mechanisms to store and search for the interaction elements. 

The $t-tuples$ could be saved in a database and then searched for later. The searching process could be enhanced by using a kind of indexing mechanism when storing them. However, these will potentially slowdown the search as there could be another outside system that it may be interfaced with. Thus, another direction is to store the elements in the same program in an array and then search for them.

%============================ Figure 7
\begin{figure} [h!]
\centering

\includegraphics [width= 0.6 \linewidth]{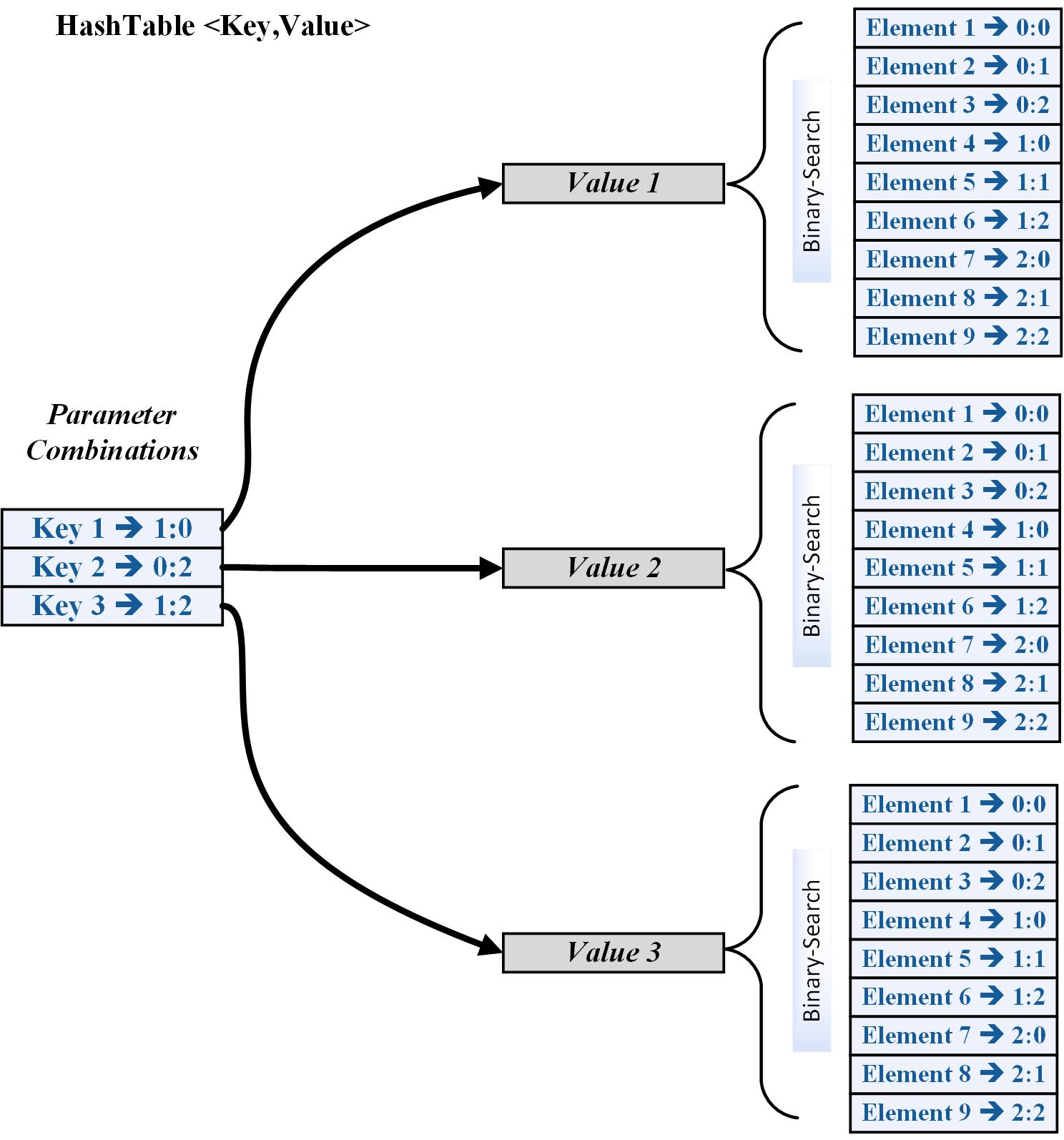}

\caption{Representation of the adopted HashTable Structure}

\label{HashTableStructure}
\end{figure}

%============================

Since there will be a huge amount of elements, the time for searching will increase dramatically with the increase in parameter and value numbers. To overcome this issue, an indexing mechanism was used to store the elements based on the interaction in a sorting array and then search for a specific element in its corresponding combination. This could speed up the search processes effectively but, the time required for finding a specific element will increase when the number of values increases (since there will be several $t-tuples$ with equal parameter combination, i.e., complexity will be $O(n)$ where $n$ is the number of stored elements). Hence, there is a need to find a new approach to store and search for the interaction elements efficiently. 

The proposed approach in this paper uses the HashTable data structure to store the interaction elements combined with a binary search process. As shown in Figure \ref{HashTableStructure}, the data structure is composed of $<Key, Value>$ pairs, and the elements are stored based on the key, with each key holding different value. When the program needs to know the number of interaction elements that could be covered by the possible solution, it will send it to the search function. The search function then searches for the interaction element in the exact $<key, value>$ pair. Finding this pair will take exactly $O(1)$ time. 

The combination and its strength will be the key for the specified element. For example, in Figure \ref{HashTableStructure}, there are three parameters in the system [0, 1 and 2]. Hence, each combination of these parameters becomes a key while its corresponding $t-tuples$ for the values will be a set of values as show in the figure. Hence, the search algorithm directly finds the key value without searching among the set of keys. Then, it searches for the specific element in the value list. To speed up this process also, a binary search is used to find the interaction element. This will allow an elimination of the complexity of the search process to $O(\log n)$ within the value "tuples" in its worst case instead of $O(n)$ within the whole $t-tuples$. Hence, without the need to search the entire $t-tuples$ set, the function knows the location of the specified element quickly and search in the tuples of the specified strength $t$.

\subsection{Generate Final Constrained Test suite}

Owing to the nature of combinatorial test generation, the original PSO algorithm cannot be used without changes and adaptation, as in case with other real world problems. Due to these necessary changes, different versions of PSO have been introduced in the literature. In this research, we also adapted and changed the original PSO algorithm to enhance the existing algorithm from the literature.

As can be seen from the original algorithm of PSO (Algorithm \ref{PSOGeneralAlgo}), the algorithm supposes an open interval search space to the continuous values of the problem’s parameters. However, in case of CIT, the values are specific in a way that each input parameter has a specific number of features. To solve a similar situation in other applications, a discrete version of PSO has been developed. We used this improvement on PSO successfully for normal CIT \cite{Mahmoud2015, Ahmed2012}.  H. Wu et. al \cite{Wu2015} also made a significant improvement to this approach to show better results. Hence, in this research, we used a discrete interval to eliminate the particles going out of the search space. As a result, we assure that the features of each parameter in the provided SUT are within our provided range.

In contrast to the implemented PSO algorithms in the literature in which the particles are represented as elements in the search space, using this algorithm for constrained CIT leads to a representation of the particles as vectors. Each particle $X_i$ in the search space is a vector with d-dimensions, in which $X_i = [d_1 , d_2 , \ldots , d_j]$ where $j$ is equal to function or parameter numbers of the SUT. Each dimension in the vector may takes from 0 to $v$ values or features.

As can be seen in the pseudo-code of the generation algorithm (Algorithm \ref{ConstraintPsoudocode}), the strategy tries to deal with the parameters and features as numbers. For example, when a system has four parameters each of them with two features, the strategy takes them in the form [2, 2, 2, 2]. In the same way, all constraints are entered into the algorithm in the same form but only those constrained functions are provided. The strategy then generates two sets of elements. The first set is the constraints that are provided to it and the second set is the all interactions among the input functions that we called $t-tuples$. The generation procedure is according to the steps provided in Section \ref{InputParameterGenerationSection}. The second set is stored in a categorised data structure with the help of hash tables as illustrated in section \ref{SearchInTheHashTableSection}. However, the first set of elements is stored in a normal array list since it is considered to be a small set and the search process does not take much time.

According to the MOPSO algorithm, the strategy generates a random search space base on the parameters and feature values of the system. Each particle (i.e., row) of this random search space then underwent an extensive evaluation by the MOPSO algorithm based on two objective functions. As illustrated in Figure \ref{OctopusStructure}, the strategy divides the search space into an equal number of particles then they were evaluated concurrently using multi-threading. The evaluation is based on the coverage of $t-tuples$ and constraint satisfaction. A best candidate is the particle that covers maximum feature combinations and satisfies a maximum number of constraints. Hence, the first objective is to cover maximum number of $t-tuples$ and the second objective is to get minimum coverage of the provided constraints. We modified the base Pareto mechanism in MOPSO algorithm by using the Pareto set as filtering mechanism. Here, if the algorithm neglects those particles with minimum violation and just chose the particles with no constraint violation, there could be an element in the neglected particle that contribute toward the best solution. Hence, we take those particles with no constraint violation or at least minimum violation during the iteration.

For each iteration, the best candidate is added to the Pareto set and the search space is updated according to the equations \ref{eq:1} and \ref{eq:2}. Instead of providing a specific number of iterations, the algorithm continues to iterate until it could not find any better solution. In this case, the best solution is chosen in the Pareto set to be added to the final test suite. The algorithm will chose the best candidate in the Pareto set with maximum coverage and $t-tuples$ and no constraint violation. 
The strategy continued in this process until all $t-tuples$ are covered. Algorithm \ref{ConstraintPsoudocode} shows a pseudo-code that illustrates the strategy step-by-step.

% ==========================================

\begin{algorithm}
\linespread{1.1}\selectfont

 \KwIn{ main configuration file}
 \KwOut{A test suite}

Get parameters $k$ and features $v$ from configuration file

Generate $t$-combinations of $k$ where $k= \lbrace k_1 , k_2 , k_3 , … k_n \rbrace $

Put $v$ for each corresponding $k$ to form $t-tuple$ 

Store $t-tuple$ set in a sorted hash table $H_t$

Store all constraints in a list $C_l$ 

Remove those tuples related to the constraints from $H_t$

Initialise $m \times k$ random swarm population $S_P$ where for $m$ particles $Xi$, where $i = {1, 2, ..., m}$

$Iter \leftarrow 1$

\While {$H_t \neq $ empty}{

    \While {$Iter <$ Max. Iter}{
    
    divide $S_P$ into equal number of particles $P_s$
    
    initialise $j$ threads where ($m$ MOD $j = 0$)
    
    \ForEach {particle in each thread of $j$}{
    
    check constraint satisfaction
    
    check coverage of $t-tuples$
    
    return best $P_s$ of each thread
    
    add best $P_s$ to Pareto set 
    
    }
    
    best $lBest \leftarrow P_s$
    
    calculate $V_i(t+1)$ according to lBest
    
   $ X_i(t) \leftarrow X_i(t+1)$ according to $V_i(t+1)$
   
    evaluate $X_i(t+1)$
    
    \If {best coverage and low constrained achieved by $lBest(t+1)$}{
    
    $lBest \leftarrow lBest(t+1)$
    }
    }
    
    Choose $P_s$ in the Pareto set with best coverage and no constraint violation
    
    $gBest \leftarrow P_s$
    
    Add $gBest$ to the test suite
    
    Remove all the related combinations from the $t-tuples$ list

}

\caption{Constrained Test Generator procedure}
\label{ConstraintPsoudocode}
\end{algorithm}

%===========================================

\begin{figure}
\centering

\includegraphics [width= 0.9\linewidth]{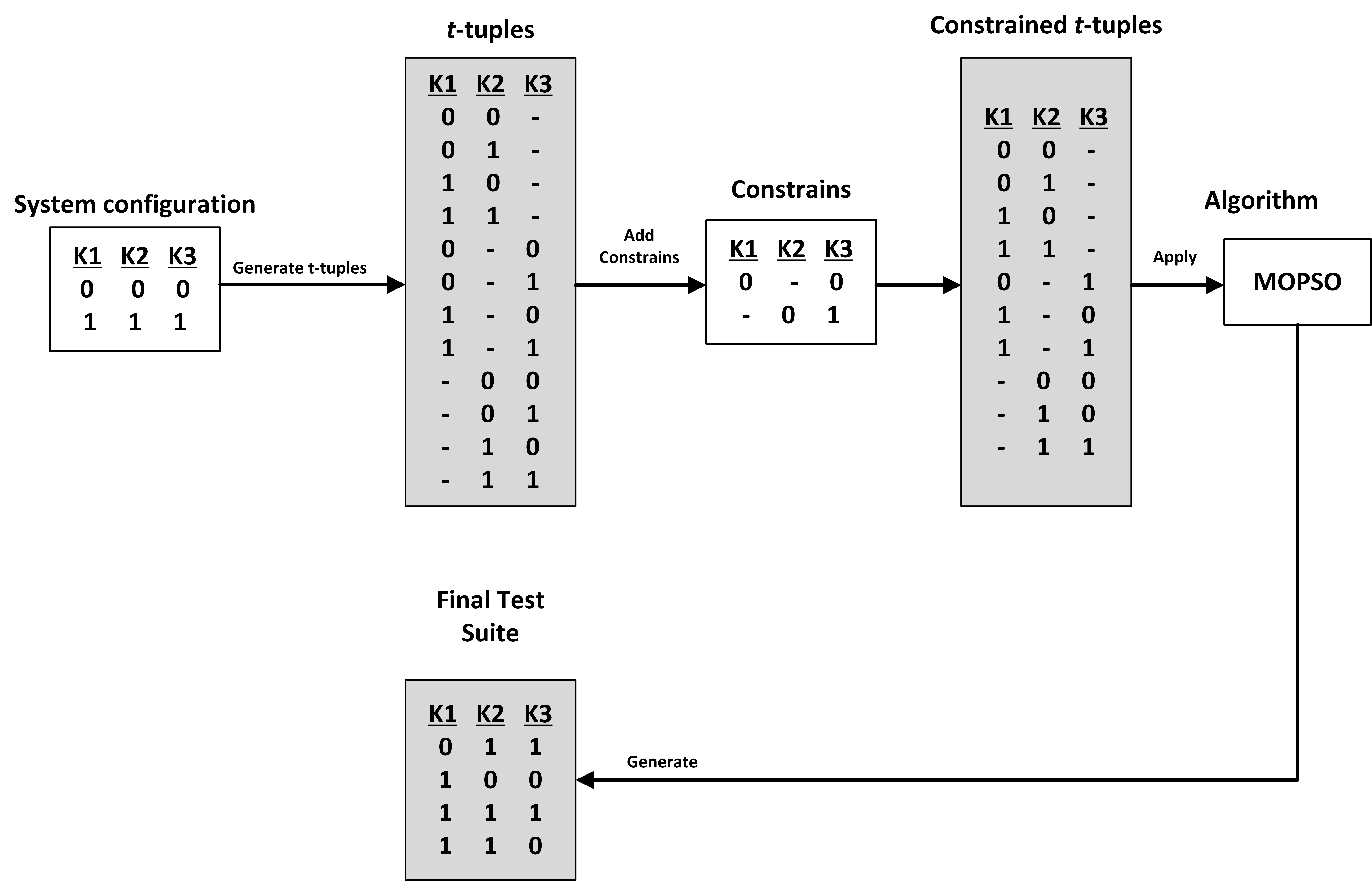}

\caption{A running example for the constrained test suite generation process}

\label{ConstraintsExampleWithTestSuiteBlockDiagram}
\end{figure}

The parameters and features’ configurations are set to the strategy by a configuration file (as shown in Algorithm \ref{ConstraintPsoudocode}). Using this configuration file, the strategy extracts the parameters and features for the SUT (Step 1). Then the t-combination of $k$ input parameters are generated based on Algorithm \ref{ParameterCombAlgo} (Step 2). To form the $t-tuple$ list that contains all the interaction elements, the strategy puts all corresponding $v$ features on each combination in the t-combination list (Step 3). These $t-tuples$ are sorted and categorised based on the combination  stored in a hash table as showed in Figure \ref{HashTableStructure} to facilitate and speed up the search process (Step 4). The input configuration file to the strategy contains also a list of constraints that will be stored in a separated list $C_l$  (Step 5). To avoid the coverage of the constrains and also to satisfy the stopping condition of the algorithm, the strategy removes those tuples related to the constraints from the t-tuples list $H_t$ (Step 6). Now, the MOPSO algorithm starts by initialising a random discrete population with dimension $m \times k$ and also initialises the interaction of the algorithm (Steps 7 and 8). The initialised population basically forms $m$ particles. The number of particles here could be subjective, however, it depends partially on the kind of the application itself. In our previous studies \cite{Ahmed2012, Bestoun2012IJICIC}, we observed that setting this number to 80 particles could contribute towards the performance and better solutions of the algorithm. The 80 particles are divided on $j$ threads (Steps 11 and 12). It should be noted that more threads could be created in the program, however, increasing the number of threads could lead to an out of memory exception when the number of parameters and features get higher. Hence, we create 8 threads here. To maintain a balance among the threads, the number of particles must be equally divided in which $m\  \textrm{mod}\ j =0$ (Step 12). Then each thread calls the checking functions for $t-tuple$ measurements and constraint satisfaction in which they return number $t-tuples$ that it can cover and also the constraints (Steps 13-16). The best solution for each thread is added then to a Pareto set (Step 17). The best solution is chosen from the Pareto set to update the search space based on it (Step 19). Then, the swarm search space is updated based on equations (1) and (2) respectively using the best solution achieved so far (Steps 20-22). The strategy will continue to run until it cannot find a better solution. In this situation, the algorithm will re-evaluate the Pareto set and choose the best solution $P_s$ with best coverage of the $t-tuples$ and no constraint violation (Step 24). This best solution will be added to the final test suite and remove its corresponding $t-tuples$ so as not to cover them again (Steps 26-27). This will stop at the end when all $t-tuples$ are covered successfully (Steps 9 and 26). For better illustration of this process, we have considered a small configuration model as a running example in Figure \ref{ConstraintsExampleWithTestSuiteBlockDiagram}. The example considered a system with three input parameters $K_1$, $K_2$ and $K_3$ each of which has two features 0 and 1. The system has two constraints: First, $K_1=0$ must not come with $K_3=0$; Second, $K_2=0$ must not come with $K_3=1$. Here, we consider interaction strength $t=2$. Clearly the figures shows how $t-tuples$ are generated and constraints handled with the $t-tuples$ set to generate the final test suite.

%============================ Figure 9

%============================

It should be mentioned here that different parameters and values of PSO are set based on our previous experience with particle swarm research. For example, 80 particles are used in the swarm which is taken from our previous experiments and research \cite{Ahmed2012}. The $C_1, C_2$, and $w$ parameters are tuned using our previous techniques using fuzzy logic process\cite{Mahmoud2015}.

%============================ Figure 9
\begin{figure}
\centering

\includegraphics [width= 0.5\linewidth]{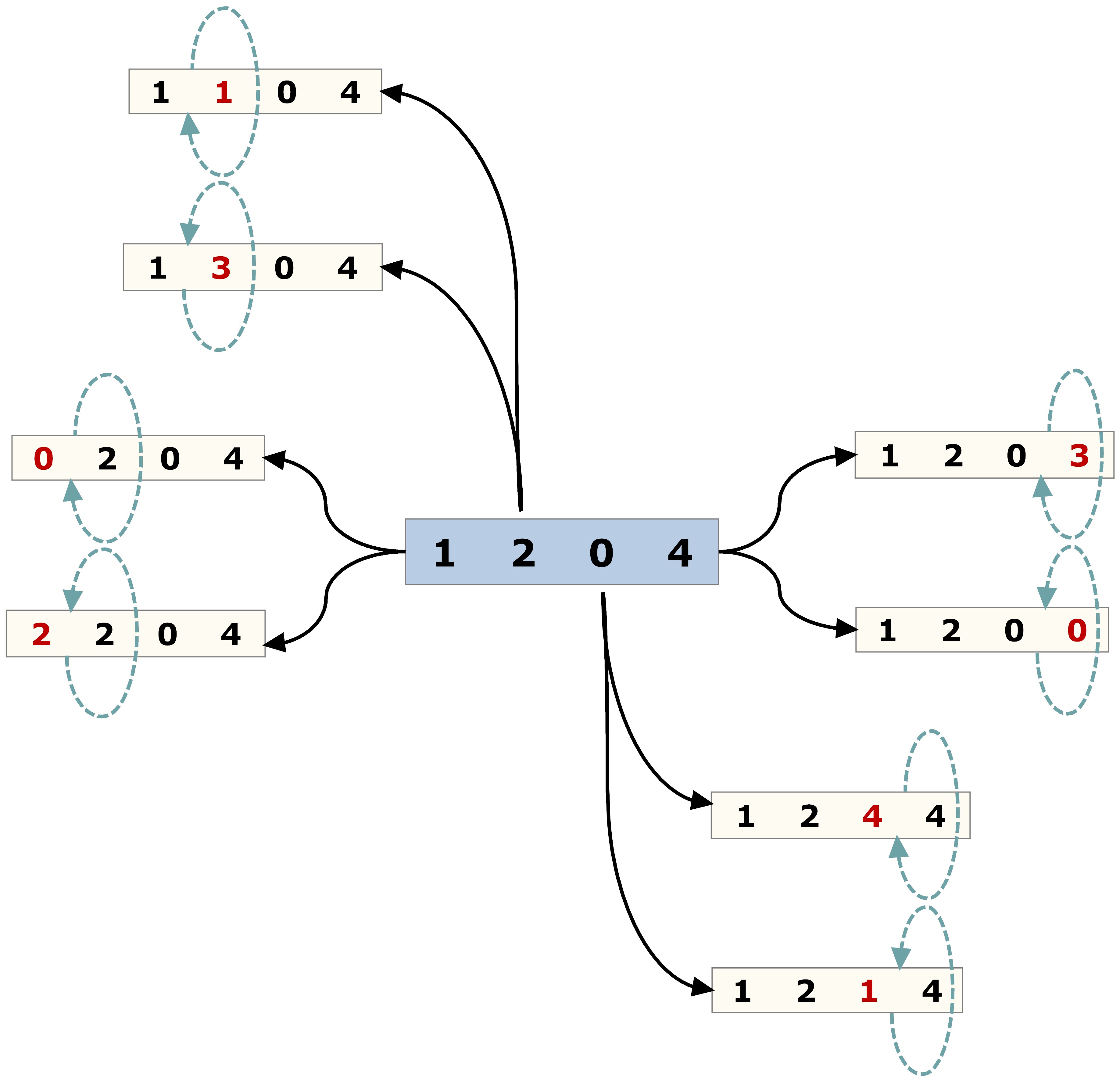}

\caption{Neighbour search process}

\label{nighbourSearchFigure}
\end{figure}

%============================

To achieve better coverage that leads to smaller size of the final test suite, we developed a technique to search in the neighbours. The algorithm first tries to find a better test case by iteration and update process, then this test case enters another short process to enhance its coverage. This will change the value of the elements to their neighbour values. Each time the modified solution is checked for coverage and constrained satisfaction. The best solution is kept until the end of the neighbour searching process. This process acts as a refinement of the local search to get better results. For better illustration, Figure \ref{nighbourSearchFigure} shows this process clearly using a running example for one solution.

\section{Evaluation Experiments} \label{EvaluationSection}

To evaluate the proposed approach, three levels of experiments were performed. The first set of experiments was to evaluate the performances of the parameter combination generation algorithm. This was carried out by running the algorithm under different conditions of input parameters and interaction degrees. In addition, other algorithms were implemented within the same environment to compare the results with them.

In the second set of experiments, the performances of the search mechanism were evaluated and different sets of benchmark were considered. The benchmarks are varied in the number of input parameters and in individual features. The performance is defined by the time in which the algorithm takes to find the set of $t-tuples$ for a specific solution. For the purpose of comparison, two other mechanisms were used in the experiments. The first mechanism stored all $t-tuples$ in an array and then searched for the elements while the second mechanism stored $t-tuples$ in an indexed array and then searched for the elements.

Finally, the third set of experiments were performed using standard benchmark experiments to evaluate the efficiency of the strategy to generate better test suites. Here we compared the strategy with other strategies that were able to generate constrained combinatorial interaction test suites. The evaluation takes the size of the generated test suite which represents the generation efficiency. For fair comparison in the case of generation time, we have implemented the available strategies for installation within our environment. 

All experiments for the available strategies as well as our strategy conducted within an environment of a desktop computer with Windows 10 installed, CPU 2.9 GHz Intel Core i5, 8 GB 1867 MHz DDR3 RAM, 512 MB of flash HDD. The strategy and its algorithms are implemented in CSharp.

\subsection{Parameter Combination Generator Evaluation}

\begin{figure} [h!]
\centering

\includegraphics [width= 0.7\linewidth]{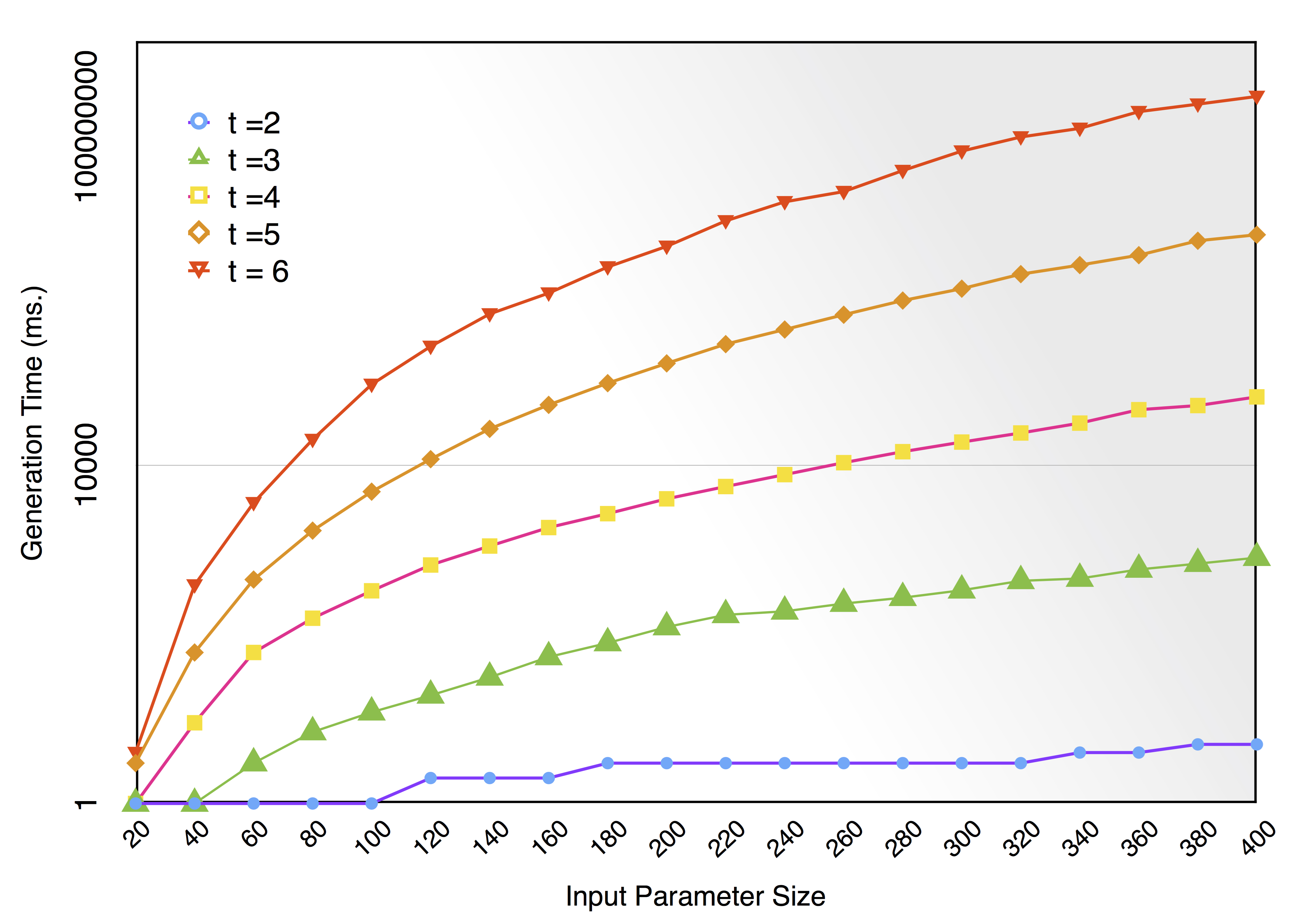}

\caption{Performance of the algorithm with the variation of input parameter and interaction strength}

\label{PerformanceParameterCombALGORITHMIC}
\end{figure}

The parameter combination generation algorithm with different parameter size interaction strength was evaluated. The parameters were varied from 20 to 400 input parameters. It is worth mentioning here that the algorithm can take more than 1000 parameters as input. However, there is no evidence in the literature showing the use of more than this number of parameter. In addition, the interaction strength was varied from 2 to 6 since this is the range of interactions used in the research so far. Figure \ref{PerformanceParameterCombALGORITHMIC} shows the comparison of these results, with the x-axis showing the parameter sizes and the y-axis showing the time in milliseconds in a logarithmic scale.  

The results showed several important points about the algorithm, and it could be noted that the algorithm performed very well for the generation. Also, it could generate the combination of 400 parameters when $t = 2$ with less than 5 milliseconds. The performance dropped when the interaction strength became higher as could be seen in the figure. However, it still performed well. For example, it could generate the combination of 100 parameters when $t = 6$ in less than 60 seconds. The drop in performance was due to the stack capacity and the several parameters pushed into it as the interaction strength increases. It could also be noted from the algorithm that when the interaction strength becomes higher, for example ($t = 6$), 6 parameters should be pushed and popped each time. This will slow down the algorithm.

%============================ Figure 10

%============================

In line with our previous discussion about the complexity of the algorithm, it could also be noted that the algorithm grow logarithmically not linearly for all values of $t$.

\subsection{Search Time Evaluation} \label{SearchTimeEvaluationSection}

\begin{figure} [h!]
\centering

\includegraphics [width= 0.7\linewidth]{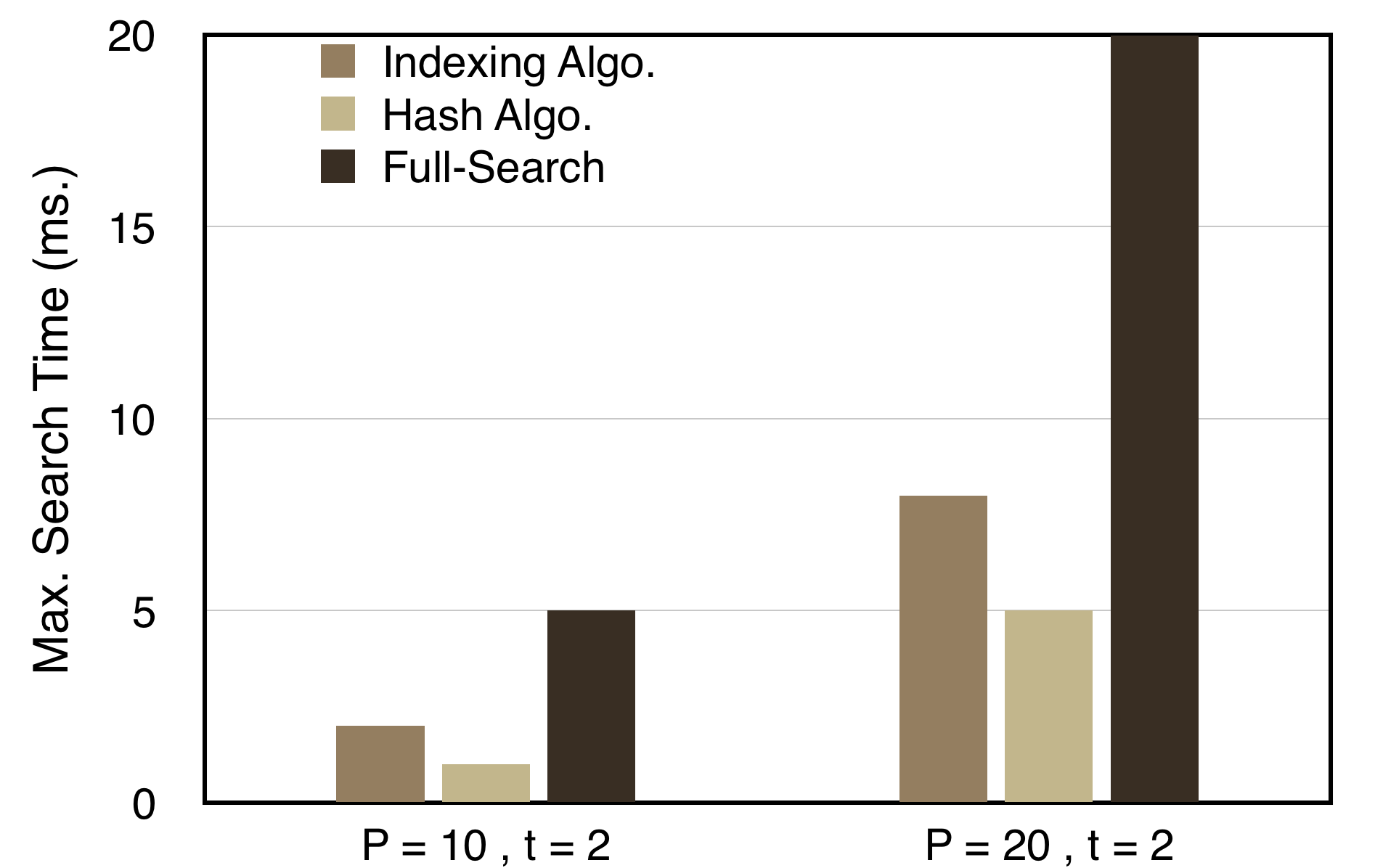}

\caption{Maximum search time measured for $t = 2$ when $v = 10$ and $P = 10$ and 20 respectively}

\label{MaximumSearchTimeT=2}
\end{figure}

\begin{figure} [h!]
\centering

\includegraphics [width= 0.7\linewidth]{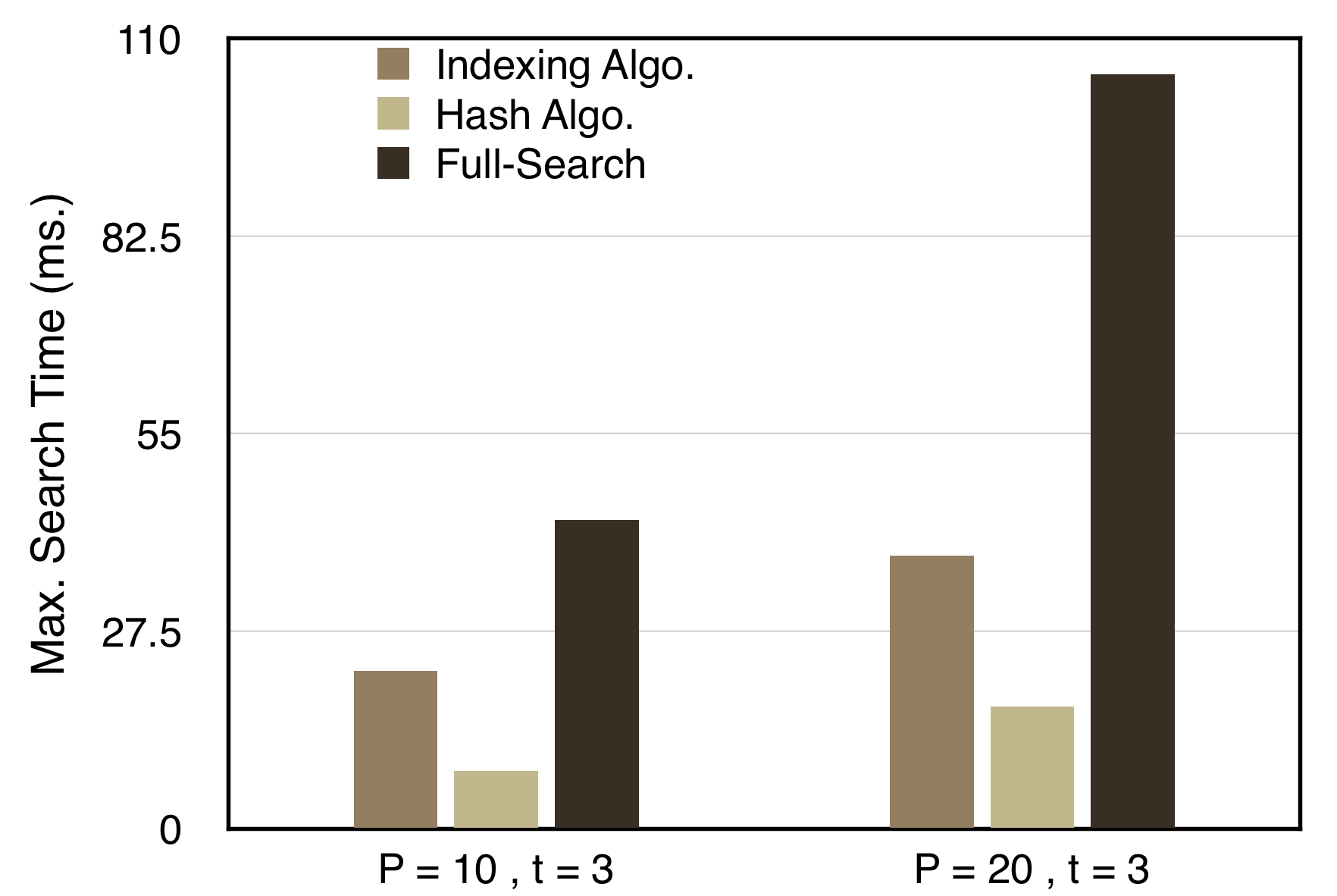}

\caption{Maximum search time measured for $t =3$ when $v =10$ and $P = 10$ and 20 respectively}

\label{MaximumSearchTimeT=3}
\end{figure}

The search time for the relevant interaction elements was measured. This time indicated the maximum time taken by the algorithm to discover the relevant $t-tuples$ for a specific solution. The maximum time was taken because the time may vary and decrease as the algorithm iterates since some of the interaction elements will be deleted. Hence, the maximum time gave a good indication about the time taken by the algorithm when the search space was full. Figures \ref{MaximumSearchTimeT=2} and \ref{MaximumSearchTimeT=3} show this time when $t = 2$ and 3 respectively for two different benchmarks.

%============================ Figure 11

%============================

As could be noted from the figures, two configurations were taken in the experiments for a covering array generation. The configurations were $CA (N; 2, 10^{10})$ and $CA (N; 2, 10^{20})$ in which the interaction strength $t =2$ then $CA (N; 3, 10^{10})$ and $CA (N; 2, 10^{20})$ with an interaction strength of $t =3$. The configurations represent perfect benchmarks for this experiment since they have many parameters and many values for the parameters. This will make the search space more complicated with many $t-tuples$.

%============================ Figure 12

%============================

As can be noticed from Figures \ref{MaximumSearchTimeT=2} and \ref{MaximumSearchTimeT=3}, the maximum search time for the interaction elements was compared for three searching mechanisms; “Hash Algo”, which is our mechanism for search with “Full-Search” and “Indexing Algo” mechanisms for searching. As mentioned earlier, the indexing mechanism saved the $t-tuples$ in a sorted array and stored the indexing of each group of $t-tuples$, however, the “Full-Search” mechanism stores all $t-tuples$ in an array without indexing and hence searches exhaustively every time in all $t-tuples$.

Both Figure \ref{MaximumSearchTimeT=2} and \ref{MaximumSearchTimeT=3} showed that our mechanism reduced the search time dramatically and hence can improve the total generation time of the solution. The figures also show that the “Full-Search” mechanism took more time to find the related interaction elements for a solution. The indexing mechanism showed better performance as compared to “Full-Search.” The searching time for the indexing mechanism was low when the number of parameters and features were low. However, as they were getting higher or the interaction strength is getting higher, the performance dropped due to the many $t-tuples$ that must be searched for in one group of indexing. Our mechanism performed better as compared to other mechanisms.

For example in Figure \ref{MaximumSearchTimeT=2}, when $P=10$ and $t=2$, the search time for our mechanism was less than 1(ms), while the indexing took more than 2 (ms). When P=20, our mechanism took less than 5(ms) for the search while indexing took more than 8(ms). This improvement in performance could be seen clearly in the case of $t=3$. When $P=10$, our mechanism took less than 8 (ms) for the search and indexing took more than 22 (ms) whereas the “Full-Search” took more than 43 (ms).

The performance of other mechanisms continue to drop in the case of $t=3$ when $P=20$ in which the indexing search time became 38 (ms) and our mechanism was 17 (ms). It should be mentioned here that this performance of the search affected the total performance cumulatively since this search process is the most consuming computation in the combinatorial search strategies.

\subsection{Generation Efficiency and Performance Evaluation}

In this section, we evaluate the generation efficiency and performance of the strategy as a whole. In CIT, the efficiency is concerned with the size of the generated test suite, whereas the performance is concerned with the generation time.   \cite{Ahmed2011, Yuan2011}. The aim of the experiment is to understand how well our strategy can generate the constrained test suites with respect to well-known benchmarks. In addition, we want to know how well the strategy performs as compared to other state-of-the-art tools and algorithms. 

For bothe efficiency and performance, we considered different benchmark problems for evaluation. The benchmarks are system models for highly configurable software. These system models are five case studies of real world configurable systems that were presented in \cite{Cohen2008} and \cite{Yu2015}. These systems are Bugzilla, Apache, GCC, SPIN simulator and SPIN verifier. Additionally, thirty configurations are generated from the characteristics of these systems. 

Bugzella \cite{BugZilla} is a defect tracker published in an open source version from Mozilla to track bugs using a special mechanism for developers. The software provides many useful features like email notification and reporting of multiple bugs. The software has a total of 52 functions in which 49 of them have binary features, one function with three features, and two functions with four features. The system model can be represented by unconstrained MCA notation as $MCA(N; t; 2^{49} 3^1 4^2)$. It can be observed from the Bugzilla specification document that it has five constraints. These constraints are four constants having two features related to each other and one constraint having three features related to each others. The same CA notation can be used here to represent these constraints as ($2^4 3^1$). As an example of constraint, the feature “sendmailnow” must be set to on when the “Postfix” feature in “Mail Transfer Agent” function is chosen.

The second case study used in this paper is Apache HTTP Server 2.2 \cite{Apache}. Apache HTTP server is an open source software used for web services that works for different platforms. To identify its various configurations and constraints among them, the software is analysed in \cite{Cohen2008}. They model the options for the software for 172 options. These options can be distributed in categories based on features for each of them and can be modelled as an unconstrained model using MCA notation, $MCA(N; t, 2^{158} 3^8 4^4 5^1 6^1)$. However, there are different constraints in the software. For example, the authenticated user must be accompanied by “AuthName” and “AuthType” directives assignment in addition to “AuthUserFile” and “AuthGroupFile.” In summary, the constraints can be modelled as ($2^3 3^1 4^2 5^1$).

GCC 4.1 \cite{GCC} is a multiple input language infrastructure that supports many languages like Java, C, C++, Fortran and many other languages. The software is also analysed by  \cite{Cohen2008} as in case of Bugzilla and Apache. GCC can be modelled as $MCA (N; t, 2^{189} 3^{10})$. In addition, there are 40 constraints among these configurations in which 37 of them have an interaction strength three and three of them have an interaction strength of two in a way that they can modelled as ($2^{37} 3^3$).

Finally, the SPIN model checker \cite{Holzmann1997} is a software tool used as a case study for evaluation. The tool is widely used as an advanced model checker. The tool can be used both as a simulator and a verifier. The simulator use of the tool simulates a single run of the system while its work as verifier is to analyse and verify all possible runs of a specific system. As a simulator, SPIN can be modelled as $MCA (N; t, 2^{13} 4^5)$ with 13 constraints of interaction strength, two that can be modelled as ($2^{13}$); while its work as an analyser can be modelled as $MCA (N; t, 2^{42} 3^2 4^{11})$, with 47 options having an iteration strength two and two binary options having an interaction strength three that can be modelled as ($2^{47} 3^2$).

In addition to these case studies, we considered another 30 systems configurations that can be derived from these case studies. The results are compared with AETG, mAETG, and SA \cite{Garvin2011}. Owing to their dependency of randomness in their meta-heuristic algorithms, the generated results are non-deterministic results. Hence, inline with the reported results by \cite{Cohen2008, Garvin2011}, each experiment is executed 50 times, (for statistical significance) then the best and average size is reported. In case of AETG and mAETG in Table \ref{SizeResultsTable}, we adopt the average sizes reported by \cite{Cohen2008} since these tools are not available for installation within our experimental environment. We did not compare the performance of the generation due to the difference in the experimental environment which leads to an unfair comparison. As mentioned earlier, the SA algorithm is implemented by Cohen et al. \cite{Cohen2008} for the generation of constrained test suites. Recently, Garvin et al. \cite{Garvin2011} implements different supported algorithms and refinements to improve the efficiency and performance of the base SA algorithm. The Implementation called Covering Array Simulated Annealing (CASA) which is available for download \cite{CASAWeb}. Considering the benchmarks, Table \ref{SizeResultsTable} shows the average reported sizes for mAETG and AETG as well as the best reported sizes for base implementation of SA compared to the best obtained results by MOPSO without considering the performance.

Table \ref{TimeResultsTable} shows the average reported sizes for the different supported algorithms for the base implementation of SA compared to MOPSO. The algorithms have been implemented within CASA tool. The average results of each algorithm is adopted from \cite{Garvin2011}. The performance evaluated and compared for CASA and MOPSO by comparing the average generation time for CASA (with all the supported algorithms) and base MOPSO. To assure fair comparison, we implement CASA within our experimental environment for the generation time.

%=================================== Table 3
\begin{table} [h!]
\centering
\includegraphics [width= 1.0\linewidth]{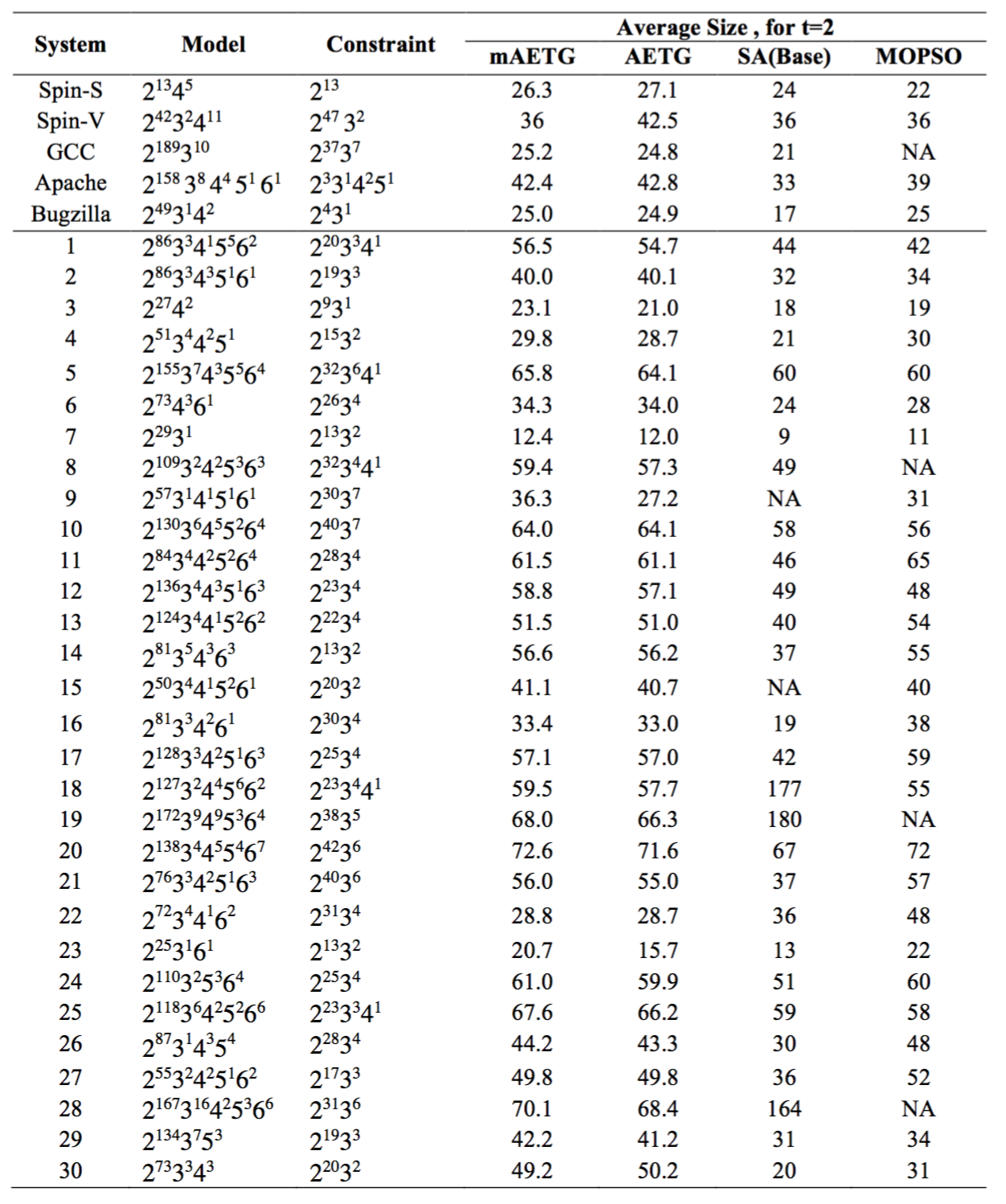}
\caption{Comparison with existing strategies for $t = 2$}
\label{SizeResultsTable}
\end{table}
% =====================================

The results in Table \ref{SizeResultsTable} shows the average sizes reported by AETG and mAETG as well as the minimum  sizes reported by base implementation of SA (i.e., SA(Base)) and MOPSO for different SUT models. The models are reported in the form of $(y_1^{k_1}  y_2 ^{k_2} . . . \ y_n ^{k_n} )$ in which for $i$ there are $k_i$ parameters that have $y_i$ features just like the notations of CAs model. In the same way, however it is denoted that for $y_i$ parameters, there are $k_i$ constraints.

It can be seen from table \ref{SizeResultsTable} that our strategy can generate small test cases in many benchmarks. Notably, the strategy can produce optimum sizes in many cases. As compared to mAETG and AETG, our strategy can produce better results. The closest competitor with our strategy is the strategy that depends on the base SA algorithm. It can be noted from the results that in those cases where our strategy is not the winner, the variation among the generated test suites in many cases is relatively modest. The efficiency of the PSO algorithms enable our strategy to outperform the AETG and mAETG since they are using the same one-row-at-time approach but with a different optimisation mechanism. However, for those cases that the SA stage is better, we observe that it is due to the way the test cases are chosen. The strategy that depends on SA does not depend on the one-row-at-time; however, it attempts to generate the entire array at a time. This leads to the generation of better results when the number of input parameter increases. This observation is in line with the observations of others which concluded that the due to the random nature of meta-heuristic algorithms, bad choice of parameters’ features could be achieved more frequently when the number of parameter grows in the case of one-parameter-at-time \cite{Cohen2003}. This interprets why SA could generate better results in case of very high parameter (probably more than 80 parameters). We believe that more improvement and research could be done for this part in the future by completing more analysis and investigations to compare these two approaches.

%=================================== Table 4
\begin{table}
\centering
\includegraphics [width= 1\linewidth]{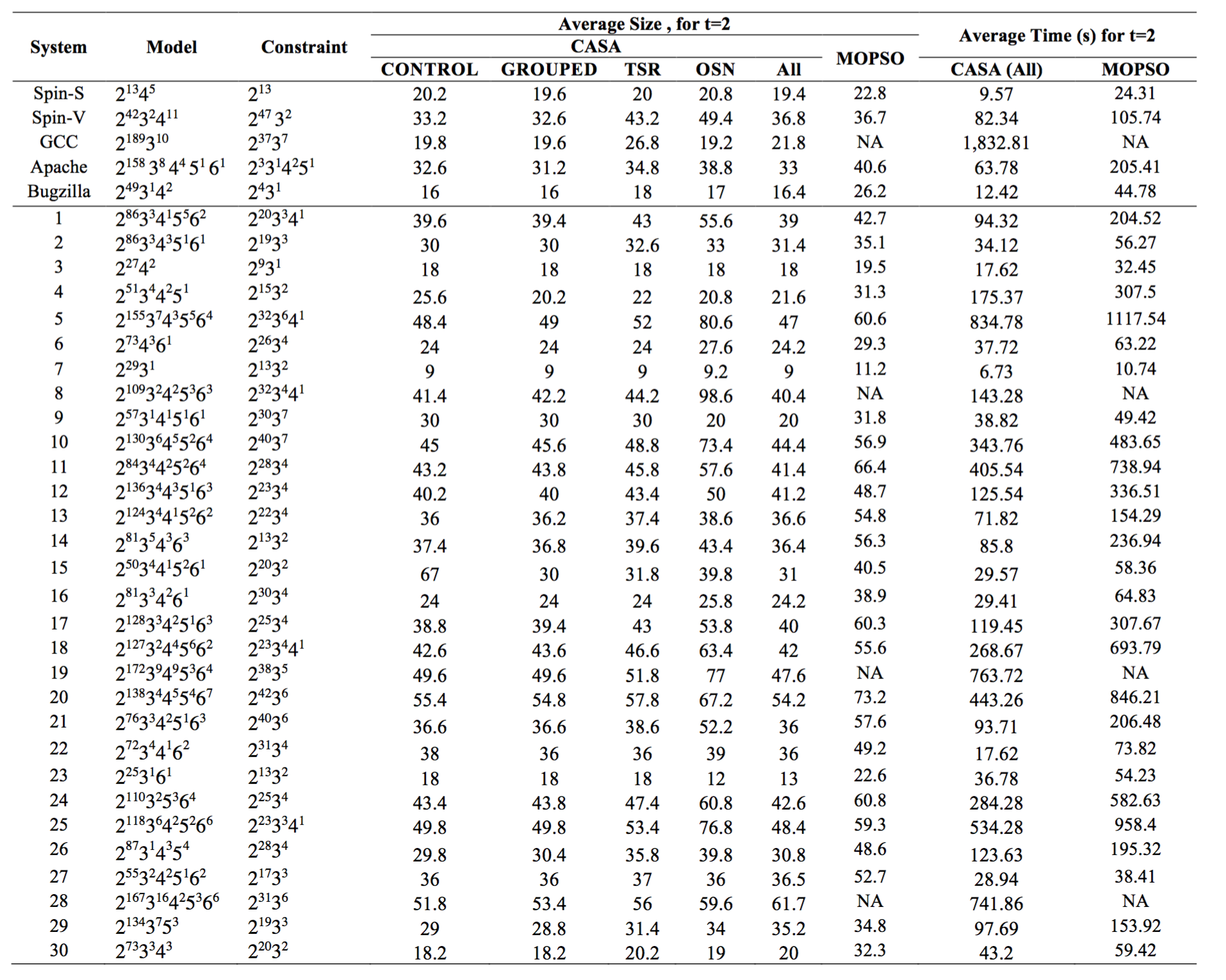}
\caption{Comparison with existing strategies for $t = 2$}
\label{TimeResultsTable}
\end{table}
% =====================================

One more observation from the results is that due to the long generation time, some of the results are marked as “NA” which indicates “Not Available results.” There could be more than one reason behind this situation. First, the strategy may need more time to cover all the $t-tuples$ and satisfy the constraints. Second, some of the configurations have more parameters and features than others that lead to many possibilities for arrangement than for better coverage of $t-tuples$. However, only few tuples will remain to be covered and the strategy takes more time to cover these tuples due to the random nature of the algorithms itself. This case is repeated also by the other strategies in some configurations.

The results in Table \ref{TimeResultsTable} shows the average sizes reported by each supported algorithm and refinements of SA compared to MOPSO. The table also reported the generation time in seconds for CASA (with all supported algorithm) and MOPSO. The results are for the aforementioned SUT models and benchmarks. The supported and refinement algorithms are CONTROL, GROUPED, TSR, OSN, and ALL. CONTROL is a version of the base algorithm refactored to support all of the suggested alterations with SAT constraint solver. GROUPED is the same as Control with code for the grouped refinements enabled. TSR builds on Grouped by adding t-set replacement. OSN extends Grouped with one-sided narrowing. ALL combines all of changes.

It can be seen from Table \ref{TimeResultsTable} that different refinement and supported algorithm for the base SA algorithm produces different results. It can be observed that the supported algorithms have different generation efficiency. The table also show us that combining all refinement for SA may not necessarily lead to better efficiency although they may contribute to improvement. Regarding the generation efficiency, the MOPSO algorithm can perform well as compared to the SA's refinement algorithms individually. It could perform well as compared to the base algorithm of SA. However, combining all the refinement and supported algorithm of SA lead to better generation efficiency. It should be mentioned that the used algorithm within our strategy is the base MOPSO algorithm. We think that the refinement and supported algorithms could be applied also to the MOPSO like using the SAT solver to support the generation of guided search space. This could be a subject for the future research. Hence, the fair comparison in case of efficiency can be achieved better in case of comparison with base SA algorithm.

Table \ref{TimeResultsTable} also shows the performance evaluation of our strategy and its comparison with CASA. It can be seen that in general our strategy perform well as compared to CASA. For both strategies, the execution time is going higher when the number of input parameters and constraints grows. The rational behind the difference in the generation time is the generation approach. We can note significant improvement in the search time for $t-tuples$ from the results presented in  \ref{SearchTimeEvaluationSection}. However, the strategy uses the one-row-at-time approach while SA generate the entire array gradually starting from smaller array. Within our strategy, there could be a few tuples that taking more time to cover. This will cause a delay in the presenting of the final test suite and may affect the total generation time.

\section{Discussion} \label{DiscussionSection}

We presented a motivating example from software product lines in Section \ref{motivatingExample} where it was shown that ordinary CIT strategies would not work in the presence of constraints. A software product line is just one example of application areas where constrained CIT is beneficial.  The other potential application area is event interaction testing for GUIs where system constraints often control triggering of events. According to Kuhn et al. \cite{Kuhn2013}, support for constraint handling is important for practical application of CIT. Our existing work supports this line of research. 

The three sets of experiments in this paper show the benefits of the implemented algorithms as well as the approach as a whole. The first set of experiments used the parameter combination generator and showed that it efficiently generates tests of different interaction strengths with number of parameters as large as 400. The second set of experiments evaluated the search time where it was shown that when the search space becomes complicated with many $t-tuples$, our implemented search mechanism outperforms full search and the indexing mechanism. The third set of experiments compared the generation strategy as a whole with mATG, AETG and SA, on a set of benchmark problems of highly configurable systems. The results showed that our approach can generate impressive results in term of test suite size in many cases as compared to the base implementation of SA. Where it was not a winner, the differences were modest at best. The results also showed that our approach can perform well in term of generation time.

We believe this paper is the first attempt to use a multi-objective meta-heuristic search approach for constrained CIT. This line of research is timely given the surge in the field of search-based software engineering \cite{Harman2007}, which is concerned with solving software engineering problems using meta-heuristic search algorithms. However, this is only a beginning of a series of empirical evaluations to be done using the multi-objective meta-heuristic search approaches for CIT.  Constraint handling in meta-heuristic search has been surveyed in  \cite{CoelloCoello2002} where several ideas could be borrowed for CIT such as integration of constraints in the fitness function as well as reducing the search space by forbidding invalid tuples. The reduction in the search space by prohibiting certain parameter-value interactions has been done using simulated annealing in \cite{Petke2013}, however the potential of multi-objective search-based CIT is yet to be exploited fully. Moreover, as the survey of CIT tools in \cite{Khalsa2014} show, not many CIT tools implement constraint-handling methods. This is even more the case when multi-objective search algorithms are used for constrained CIT. Building on the work in this paper, our ultimate goal is to develop a stable tool where one or more multi-objective search-based algorithms are implemented for constrained CIT.

\section{Threats to Validity} \label{ThreadsSection}

As in the case with other research, this research faces some threats to validity. We have tried to focus and eliminate these threats as much as we can; however, some of these threats are out of our control. The first threat is the lack of results for other strategies. Within CIT, there could be many strategies to be compared with. However, in the case of constrained CIT, there are few generation strategies since it is new direction in CIT. We found those mentioned strategies and benchmarks for comparison which are the best results published so far. However, there could be many other benchmarks in which such strategies might not be efficient.

Another important threat is the time of generation for each strategy. It is well-known that the size of the test suite is not affected by the implemented environment; however, the time of generation for the test suite is highly subjective to the running environment. Due to this reason, we cannot compare the time of generation directly with those published results in the literature. Hence, to make a fair comparison with the time of generation, it is compulsory to have all strategies implemented within the same environment which is difficult in our case due to unavailability of those compared strategies for public use. We compared the generation time of our strategy with CASA tool that contain the SA implementation. The comparison could give an indication for the generation time; however there are two threats here. First, the implementation language differences. CASA implemented using C++ while our strategy implemented using CSharp. Second, the implemented algorithm inside CASA is not the base SA algorithm. The strategy contains different refinement and supporting algorithms that leads to improve the final result. These techniques and others could be applied within our strategy to improve the final results. For example, we can add a compaction algorithm in order to compact the final result before display to further merge the test cases to get optimum results. However, we are more interested to the pure performance of MOPSO. Hence, fair comparison can be achieved when we compare MOPSO with base implementation of SA.

Finally, the number of constraints specified for the program could form another source of threat. The benchmark experiments used in this research are models for real world software systems. As we can see, there are few constraints specified for each system. The program generates the final test suites within short time.  In case of high constraints' number, we expect longer time for generation due to the randomness nature of the PSO algorithm. However, this case is not applicable until now in real-world application testing.

\section{Conclusions} \label{ConclusionSection}

In this paper, we have presented our new approach to generate constrained combinatorial interaction test suites. The approach has been implemented successfully in a strategy. The strategy takes the input parameters and features of each of them as an input configuration. These configurations contain different constraints that must be provided for the strategy. The output of the strategy is an optimal combinatorial interaction test suite that satisfies the constraints among the combinations of the input features. To generate this test suite, different, new, and effective algorithms have been implemented within the strategy. A new mechanism is used to speed up the generation and search for interactions by designing and using efficient data structures. Using our experience with CIT strategies, multi-objective PSO is used to find an optimal solution for the final test suites. The strategy operates the algorithms in parallel by using multi-threading.

In order to conduct an extensive evaluation, different experiments have been applied. Experiments are conducted to evaluate each algorithm inside the strategy and also, the strategy as a whole. To do so, well-known benchmarks have been selected in different stages of the experiments in addition to other experiments that were designed for this purpose. In addition to the comparison with this published results of other strategies, the strategy is compared with other available strategies by implementing them within our environment or adopting the published results. Experiment results showed that the developed algorithm performed well under circumstances of benchmarks. The experiments also showed that a new strategy can produce impressive results as compared to other available strategies with binomial time for generation.

The strategy could be applicable for testing in different software disciplines. For example, the strategy could be applied for event interactions testing for graphical user interface (GUI) when there are different constraints among the event interactions. The strategy is highly applicable for software product line testing to find and test different products. In this case, there could be many interaction constraints among the features of the products.

\section{Acknowledgment}
The first author of the paper would like to thank IDSIA institute and Swiss Excellence Scholarship for hosting and supporting this research. Kamal Z. Zamli is funded by the Science Fund Grant RDU140503- “Constraints t-way Testing Strategy with Modified Condition Decision Coverage” from the Ministry of Science, Technology, and Innovation (MOSTI), Malaysia.  Wasif Afzal is funded by The Knowledge Foundation (KKS) through the project 20130085: Testing of Critical System Characteristics (TOCSYC).

\section*{References}

\bibliography{mybibfile}

\end{document}